\documentclass[final,1p,sort&compress]{elsarticle}
\usepackage{amsmath,amssymb,amscd}
\usepackage[notref,notcite]{showkeys}
\journal{Nuclear Physics B}
\newcommand{\al}{\alpha}
\newcommand{\be}{\beta}
\newcommand{\de}{\delta}
\newcommand{\ep}{\epsilon}

\newcommand{\ga}{\gamma}

\newcommand{\la}{\lambda}

\newcommand{\si}{\sigma}
\newcommand{\te}{\theta}
\newcommand{\vt}{\vartheta}
\newcommand{\vp}{\varphi}
\newcommand{\ze}{\zeta}
\newcommand{\La}{\Lambda}
\newcommand{\Si}{\Sigma}
\newcommand{\ba}{\mathbf{a}}

\newcommand{\bev}{\mathbf{e}}
\newcommand{\bk}{\mathbf{k}}

\newcommand{\bn}{\mathbf{n}}
\newcommand{\bp}{\mathbf{p}}
\newcommand{\bs}{\mathbf{s}}
\newcommand{\bx}{\mathbf{x}}

\newcommand{\bnu}{{\boldsymbol{\nu}}}

\newcommand{\bt}{{\boldsymbol{\vt}}}
\newcommand{\bte}{{\boldsymbol{\te}}}
\newcommand{\bu}{\mathbf{u}}

\newcommand{\by}{\mathbf{y}}
\newcommand{\bz}{\mathbf{z}}

\newcommand{\tH}{\tilde{H}}
\newcommand{\tK}{\widetilde{K}}

\newcommand{\tS}{\tilde{S}}

\newcommand{\NN}{{\mathbb N}}
\newcommand{\RR}{{\mathbb R}}
\newcommand{\CC}{{\mathbb C}}
\newcommand{\ZZ}{{\mathbb Z}}
\newcommand{\cE}{{\mathcal E}}

\newcommand{\cH}{{\mathcal H}}

\newcommand{\cP}{{\mathcal P}}
\newcommand{\cQ}{{\mathcal Q}}

\newcommand{\cZ}{{\mathcal Z}}
\newcommand{\fH}{{\mathfrak H}}

\newcommand{\fT}{{\mathfrak T}}
\newcommand{\fW}{{\mathfrak W}}
\newcommand\BA{\,\overline{\!A}{}}
\newcommand\BC{\,\overline{\!C}{}}

\newcommand\Esc{E^{\mathrm{sc}}}
\newcommand\Lsc{\La_{\mathrm{sc}}}
\newcommand\Hsc{H_{\mathrm{sc}}}
\newcommand\tHsc{\tH_{\mathrm{sc}}}
\newcommand\Zsc{Z_{\mathrm{sc}}}
\newcommand\fHsc{\fH_{\mathrm{sc}}}
\newcommand\Lasc{\La_{\mathrm{sc}}}
\newcommand\fWa{\mathfrak{W}_{\mathrm{a}}}
\newcommand{\pa}{\partial}

\newcommand{\id}{1\hspace{-.25em}{\rm I}}

\newcommand\ket[1]{|#1\rangle}

\newcommand{\ms}{\mspace{1mu}}
\renewcommand{\le}{\leqslant}
\renewcommand{\ge}{\geqslant}

\renewcommand{\geq}{\geqslant}

\newcommand{\tr}{\operatorname{tr}}

\newcommand{\card}{\operatorname{card}}

\newcommand{\iu}{\mathrm{i}}
\newcommand{\e}{\mathrm{e}}

\newdefinition{remark}{Remark}
\begin{document}
\begin{frontmatter}
  \title{The spin Sutherland model of $D_N$ type\\and its associated spin chain}
  \author[BBM]{B. Basu-Mallick}

  \author[UCM]{F. Finkel}

  \author[UCM]{A. Gonz\'alez-L\'opez\corref{cor}}
  \ead{artemio@fis.ucm.es}
  
  \cortext[cor]{Corresponding author}

  \address[BBM]{Theory Group, Saha Institute of Nuclear Physics, 1/AF Bidhan Nagar, Kolkata 700
    064, India}
  \address[UCM]{Departamento de F\'\i sica Te\'orica II, Universidad Complutense, 28040
    Madrid, Spain}

  \date{August 26, 2010}
  \begin{abstract}
    In this paper we study the $\mathrm{su}(m)$ spin Sutherland (trigonometric) model of $D_N$
    type and its related spin chain of Haldane--Shastry type obtained by means of Polychronakos's
    freezing trick. As in the rational case recently studied by the authors, we show that these
    are new models, whose properties cannot be simply deduced from those of their well-known
    $BC_N$ counterparts by taking a suitable limit. We identify the Weyl-invariant extended
    configuration space of the spin dynamical model, which turns out to be the $N$-dimensional
    generalization of a rhombic dodecahedron. This is in fact one of the reasons underlying the
    greater complexity of the models studied in this paper in comparison with both
    their rational and $BC_N$ counterparts.
    By constructing a non-orthogonal basis of the Hilbert space of the spin
    dynamical model on which its Hamiltonian acts triangularly, we
    compute its spectrum in closed form. Using this result and applying the
    freezing trick, we derive an exact expression for the partition function
    of the associated Haldane--Shastry spin chain of $D_N$ type.
  \end{abstract}
\end{frontmatter}

\section{Introduction}\label{sec.intro}
Recent studies have revealed that exactly solvable and integrable one-di\-men\-sion\-al quantum many
body systems with long-range interactions~\cite{Ca71,Su71,Su72, OP83, Ha88,Sh88,Po93,Ha96} are
closely connected with a wide range of topics in modern physics as well as mathematics. In
particular, this type of exactly solvable systems have appeared as prototype models of various
condensed matter systems exhibiting generalized exclusion statistics~\cite{Ha96,MS94,Po06},
quantum Hall effect~\cite{AI94} and quantum electric transport phenomena~\cite{BR94,Ca95}. In the
context of high-energy physics, the dynamics of particles or fields in the near-horizon region of
black holes has been described through such integrable systems~\cite{CDKKTP98,GSV00,GT99}. More
recently, quantum integrable spin chains with long-range interaction have played a key role in
calculating higher loop effects in the spectra of trace operators of planar ${\mathcal N}=4$ super
Yang--Mills theory~\cite{BKS03,Be04,BBL09}. Furthermore, this type of quantum integrable systems
are found to be connected with different areas of mathematics like random matrix
theory~\cite{TSA95}, multivariate orthogonal polynomials~\cite{Fo94,BF97,Di97,UW97}, Dunkl
operators~\cite{Du98,FGGRZ01}, and Yangian quantum groups~\cite{BGHP93,Hi95npb,Ba99,BE08}.

Due to such a large variety of potential applications, the construction of new quantum integrable
systems with long-range interactions and the computation of their exact solutions have emerged as
an important area of activity in the current literature. The study of this type of systems with
dynamical degrees of freedom was pioneered by Calogero, who found the exact spectrum of an
$N$-particle quantum system on the line with two-body interactions inversely proportional to the
square of the distances and subject to a confining harmonic potential~\cite{Ca71}. An exactly
solvable trigonometric variant of this rational Calogero model was subsequently proposed by
Sutherland~\cite{Su71,Su72}. The particles in this so-called Sutherland model move on a circle,
with two-body interactions proportional to the inverse square of their chord distances. In a
parallel development, Haldane and Shastry found an exactly solvable quantum spin-$\frac{1}{2}$
chain with long-range interactions~\cite{Ha88,Sh88}. The lattice sites of this su$(2)$
Haldane--Shastry (HS) spin chain are equally spaced on a circle, all spins interacting with one
another through pairwise exchange interactions inversely proportional to the square of their chord
distances. A close relation between the HS chain with su($m$) spin degrees of freedom and the
su($m$) spin generalization of the Sutherland model~\cite{HH92,HW93,MP93} was subsequently
established by using the so-called ``freezing trick''~\cite{Po93,SS93}. More precisely, it is
found that in the strong coupling limit the particles in the spin Sutherland model ``freeze'' at
the coordinates of the equilibrium position of the scalar part of the potential, and the dynamical
and spin degrees of freedom decouple. The equilibrium coordinates coincide with the equally spaced
lattice points of the HS spin chain, so that the decoupled spin degrees of freedom are governed by
the Hamiltonian of the su($m$) HS model. Moreover, in this freezing limit the conserved quantities
of the spin Sutherland model immediately yield those of the HS spin chain, thereby explaining its
complete integrability. Application of this freezing trick to the rational Calogero model with
spin degrees of freedom leads to a new integrable spin chain with long-range
interaction~\cite{Po93}. The sites of this chain ---commonly known in the literature as the
Polychronakos or Polychronakos--Frahm (PF) spin chain--- are unequally spaced on a line, and in
fact coincide with the zeros of the Hermite polynomial of degree $N$~\cite{Fr93}. By applying the
freezing trick, the exact partition functions of the PF and HS spin chains have also been exactly
computed~\cite{Po94,FG05}.

The above mentioned type of quantum integrable systems can be generalized to form a much wider
class by taking advantage of their hidden mathematical structure. Indeed, Olshanetsky and
Perelomov established the existence of an underlying $A_N$ root system structure for both the
spinless Calogero and Sutherland models, and constructed generalizations thereof associated with
any (extended) root system~\cite{OP83}. Spin generalizations of the $BC_N$ Calogero and Sutherland
models have also been proposed, and various properties of the related lattice models of HS type
have been studied with the help of the freezing trick~\cite{BPS95, Ya95, YT96, CS02, EFGR05,
  BFGR08}. Among the other classical root systems, the exceptional ones are comparatively less
interesting, since their associated models consist of at most $8$ particles. Until recently, the
$B_N$, $C_N$ and $D_N$ Calogero--Sutherland models (particularly the corresponding spin models)
have been largely ignored, probably due to the fact that they were believed to be simple limiting
cases of their $BC_N$ counterparts. However, in a recent paper~\cite{BFG09} the present authors
have computed the spectrum of the su($m$) spin Calogero model of $D_N$ type, thereby showing that
this model is in fact a singular limit of its $BC_N$ version. More precisely, it is well known
that the Hilbert space of the spin Calogero model associated with the $BC_N$ root system can be
constructed from the Hilbert space of an auxiliary differential-difference operator by using a
single projector. In contrast, it is found that two independent projectors of $BC_N$ type with
opposite ``chiralities'' are needed to construct the Hilbert space of the $D_N$-type spin Calogero
model from that of the corresponding auxiliary operator~\cite{BFG09}. Consequently, the Hilbert
space of the latter model can be expressed as a direct sum of the Hilbert spaces associated with
two different $BC_N$ models with opposite chiralities. This explains why the spectrum of the $D_N$
model cannot be obtained as the limit of its $BC_N$ counterpart when one of the coupling constants
tends to zero. In Ref.~\cite{BFG09} we also studied the spin chain associated with the $D_N$-type
spin Calogero model, showing that its Hamiltonian differs from the limit of its $BC_N$ analog by a
term which can be interpreted as an impurity interaction at one end of the chain. By applying
Polychronakos's freezing trick we were also able to compute the chain's partition function in
closed form, showing again that its spectrum markedly differs from that of its $BC_N$ counterpart.

In this paper we study the trigonometric variant of the $D_N$-type spin Calogero model and its
freezing limit, i.e., the $D_N$-type spin Sutherland model and its related spin chain. Just as in
the rational case, the Hamiltonian of the $D_N$-type spin Sutherland model can be formally
obtained as a certain limit of its $BC_N$ counterpart. However, the relation between these models
turns out to be even more subtle than in the rational case. Roughly speaking, this is due to the
fact that the Weyl-invariant extended configuration space of the $D_N$ model ---which turns out to
be the $N$-dimensional generalization of a rhombic dodecahedron--- does not coincide with that of
the $BC_N$ model, which is simply a hypercube. As a consequence, the (scaled) Fourier basis of the
Hilbert space of the $BC_N$ model's auxiliary operator no longer spans a complete set of the
Hilbert space of the auxiliary operator of the $D_N$ model. This entails an additional level of
difficulty (but also of interest) by comparison with the rational case, for which the auxiliary
operators of the $BC_N$ and $D_N$ models share the same Hilbert space. On the other hand, as in
the rational case, we shall still need two projectors of $BC_N$ type with opposite chiralities in
order to construct the Hilbert space of the $D_N$ spin model from that of its corresponding
auxiliary operator. Therefore, the Hilbert space of the $D_N$ spin model actually consists of {\em
  four} ---and not two, as in the rational case--- different sectors, characterized by their
chirality and parity under reflections of the particles' coordinates. This fundamental difference
explains why the spectrum of the $D_N$-type spin Sutherland model is essentially different from
that of its $BC_N$ counterpart. It also accounts for the greater complexity of the partition
function of the associated chain of $D_N$ type (which we have also computed in closed form by
means of the freezing trick) compared to its $BC_N$ version studied in Ref.~\cite{EFGR05}.
  
The paper is organized as follows. In Section~\ref{sec.themodels} we introduce the Hamiltonians
$H$, $\Hsc$ and $\cH$ of the $D_N$-type spin Sutherland model, its scalar version, and the
associated spin chain of HS type, respectively. We show that the sites of this chain, defined as
the coordinates of the (unique) equilibrium point of the scalar part of the spin Hamiltonian in
the principal Weyl alcove of the $D_N$ root system, can be expressed in terms of the roots of a
suitable Jacobi polynomial. Using this characterization, we prove that the Hamiltonian $\cH$
differs from the limit of its $BC_N$ counterpart by a spin reversing term at each end of the
chain.

Section~\ref{sec.spectrum} is devoted to the computation of the spectrum of the Hamiltonians $H$
and $\Hsc$ using an auxiliary {\em scalar} differential-difference operator~$H'$. In order to
improve the clarity of the exposition, we have divided it into four subsections. In the first one,
we show that for any one-dimensional representation $\pi$ of the $D_N$ Weyl group $\fW$, the
Hamiltonians $H$ and $\Hsc$ are equivalent (isospectral) to their $\pi$-symmetric extensions to
the $\fW$-orbit $C$ of the configuration space. The representation $\pi$ is then uniquely
determined by requiring that the action of the extension of $H$ (resp.~$\Hsc$) coincide with that
of $H'\otimes\id$ (resp.~$H'$) on the subspace of $L^2(C)\otimes\Si$ (resp.~$L^2(C)$) of
$\pi$-symmetric functions, where $\Si$ denotes the su($m$) spin space. This property, together
with the fact that $H'$ commutes with the projector onto $\pi$-symmetric functions, enable us to
evaluate the spectra of $H$ and $\Hsc$ by triangularizing the simpler operator $H'$. The first
step in this direction is to construct a (non-orthogonal) basis of $L^2(C)$, which includes as a
proper subset the limit of the basis of $L^2\big([-\frac\pi2,\frac\pi2]\big)$ for the $BC_N$
model. As a by-product of this construction, we show that the translations of $C$ generate a
tessellation of the $N$-dimensional Euclidean space. By expressing $H'$ as a sum of squares of a
suitable family of Dunkl operators, we show that when the above basis of $L^2(C)$ is ordered in an
appropriate way the action of $H'$ becomes triangular. Using this basis, in the last subsection we
construct a corresponding (non-orthogonal) basis of the subspace of $\pi$-symmetric functions
$L^2(C)\otimes\Si$ (resp.~$L^2(C)$) in which the action of $H$ (resp.~$\Hsc$) is also triangular.
This completes the calculation of the spectra of the Hamiltonians $H$ and $\Hsc$.

In Section~\ref{sec.pf} we compute the partition function $\cZ$ of the spin chain $\cH$ using
Polychronakos's freezing trick. More precisely, we evaluate the partition function $\cZ$ as the
large coupling constant limit of the quotient of the partition functions of $H$ and $\Hsc$. The
resulting formula exhibits a greater complexity than its rational counterpart and, in particular,
cannot be expressed in a simple way in terms of the partition functions of the $BC_N$
trigonometric spin chains. We end up this section by showing that the latter expression for the
partition function is indeed a polynomial in $q\equiv\e^{-1/(k_{\mathrm B}T)}$, as should be the
case for a finite system with nonnegative integer energies. Finally, a brief summary of the
paper's results is presented in Section~\ref{sec.remarks}.

\section{The models}\label{sec.themodels}
Our starting point will be a brief review of the su($m$) spin Sutherland model of $BC_N$ type,
with Hamiltonian~\cite{Ya95,EFGR05}
\begin{align}
  H^{(\mathrm B)} = -\sum_i \pa_{x_i}^2 &+a\,\sum_{i\neq j}\big[\sin^{-2}
  x_{ij}^-\,(a+\,S_{ij})+\sin^{-2} x_{ij}^+\,(a+\tS_{ij})\big]\notag\\
  &+b\,\sum_i \sin^{-2}\!x_i\,(b-\ep S_i)+b'\,\sum_i \cos^{-2}\!x_i\,\big(b'-\ep S_i\big)\,.
  \label{HSB}
\end{align}
Here the sums run from $1$ to $N$ (as always hereafter, unless otherwise stated), $a,b,b'>1/2$,
$\epsilon =\pm 1$, $x_{ij}^{\pm}=x_i\pm x_j$. The operators $S_{ij}$ and $S_i$ in the latter equation
act on the finite-dimensional Hilbert space
\begin{equation}
  \label{spinbasis}
  \Si=\Big\langle\,|s_1,\dots,s_N\rangle\;\big|\; s_i=-M,-M+1,\dots,M\Big\rangle,
  \qquad M\equiv\frac{m-1}2\in\frac\NN2\,,
\end{equation}
associated to the particles' internal degrees of freedom, as follows:
\begin{equation}
  \begin{aligned}
    & S_{ij}|s_1,\dots,s_i,\dots,s_j,\dots,s_N\rangle=|s_1,\dots,
    s_j,\dots,s_i,\dots,s_N\rangle\,,\\
    & S_i|s_1,\dots,s_i,\dots,s_N\rangle=|s_1,\dots,-s_i,\dots,s_N\rangle\,.
    \label{SS}
  \end{aligned}
\end{equation}
We have also used the customary notation $\tS_{ij} =S_iS_jS_{ij}$. Note that the spin operators
$S_{ij}$ and $S_i$ can be expressed in terms of the fundamental su($m$) spin generators $J^\al_k$
at the site $k$ (with the normalization $\tr(J^\al_kJ^{\ga}_k)=\frac12\de^{\al\ga}$) as
\[
S_{ij}=\frac1m+2\sum_{\al=1}^{m^2-1}J_i^\al J_j^\al\,,\qquad S_i=\sqrt{2m}\,J^1_i\,.
\]
Due to the singularities at the hyperplanes $x_i\pm x_j=k\pi$, $x_i=k\pi$ and $x_i=\frac\pi2+k\pi$
(with $1\le i<j\le N$ and $k\in\ZZ$), the configuration space of the Sutherland
Hamiltonian~\eqref{HSB} can be taken as the principal Weyl alcove
\begin{equation}\label{CBHS}
  A^{(\mathrm B)} = \Big\{\bx\in\RR^N: 0<x_1<x_2<\cdots<x_N<\frac\pi2\Big\}
\end{equation}
of the $BC_N$ root system. The spectrum of the spin model~\eqref{HSB} was computed in
Ref.~\cite{EFGR05} by constructing a suitable basis of its Hilbert space in which the Hamiltonian
$H^{(\mathrm B)}$ is represented by a triangular matrix.

Applying the so-called freezing trick~\cite{Po93} to the Hamiltonian~\eqref{HSB} with $b=\be a$
and $b'=\be' a$ one obtains the su($m$) Haldane--Shastry (antiferromagnetic) spin chain of $BC_N$
type, whose Hamiltonian we shall take as
\begin{multline}
  \label{HSchain}
  \cH^{(\mathrm B)}=\frac12\,\sum_{i<j}\Big[\sin^{-2}\te_{ij}^-\,(1+S_{ij})
  +\sin^{-2}\te_{ij}^+\,(1+\tS_{ij})\Big]\\
  {}+\frac14\,\sum_i\big(\be\,\sin^{-2}\te_i+\be'\,\cos^{-2}\te_i\big)(1-\ep S_i)\,.
\end{multline}
Here $\te_{ij}^\pm=\te_i\pm\te_j$, where $\bte=(\te_1,\ldots,\te_N)$ is the unique
equilibrium~\cite{CS02} in the set $A^{(\mathrm B)}$ of the scalar potential
\begin{equation}
  \label{UB}
  U^{(\mathrm B)}(\bx)=\sum_{i\neq
    j}\big(\sin^{-2}x_{ij}^-+\sin^{-2}x_{ij}^+\big)
  +\sum_i\big(\be^2\sin^{-2}x_i+\be'^2\cos^{-2}x_i\big)\,.
\end{equation}
As shown in the latter reference, the lattice sites $\te_i$ are related to the zeros $\ze_i$ of
the Jacobi polynomial $P_N^{(\be-1,\be'-1)}$ by
\begin{equation}\label{zi}
  \ze_i=\cos(2\te_i)\,.
\end{equation}
The spin chain~\eqref{HSchain} was studied in Ref.~\cite{EFGR05}, where its partition function was
computed in closed form with the help of the freezing trick.\medskip

The Hamiltonian $H$ of the su($m$) spin Sutherland model of $D_N$ type is defined by setting
$b=b'=0$ in Eq.~\eqref{HSB}, i.e.,
\begin{equation}
  H = -\sum_i \pa_{x_i}^2 + a\,\sum_{i\neq j}\big[\sin^{-2}
  x_{ij}^-\,(a+\,S_{ij})+\sin^{-2} x_{ij}^+\,(a+\tS_{ij})\big]\,.
  \label{H}
\end{equation}
The configuration space $A$ of the $D_N$ model~\eqref{H} is determined by the hard-core
singularities of the Hamiltonian on the hyperplanes $x_i\pm x_j=k\pi$ (with $i\ne j$ and $k\in\ZZ$).
More precisely, we shall take as $A$ the open subset of $\RR^N$ defined by the inequalities
\begin{equation}\label{defA}
  0<x_i\pm x_j<\pi\,,\qquad 1\le j<i\le N.
\end{equation}
If $N>2$ (as we shall assume hereafter), it is straightforward to check that this set can be
equivalently expressed as
\begin{equation}
  \label{CDHS}
  A=\{\bx\in\RR^N:|x_1|<x_2<\cdots<x_N<\pi-x_{N-1}\}\,,
\end{equation}
which is again the principal Weyl alcove of the $D_N$ root system
\begin{equation}\label{DNroots}
  \frac1\pi\,(\pm \bev_i\pm \bev_j)\,,\qquad 1\le i<j\le N\,.
\end{equation}
The points $\bx\in A$ clearly satisfy the inequalities
\begin{equation}\label{ineqs}
  0<x_2<\cdots<x_{N-1}<\pi/2,\quad x_1>-\pi/2,\quad x_N<\pi\,;
\end{equation}
note, in particular, that the set $A$ properly contains the $BC_N$ configuration
space~\eqref{CBHS}.

Similarly (cf.~\eqref{HSchain} and~\eqref{UB}), we define the Hamiltonian of the su($m$) HS spin
chain of $D_N$ type as
\begin{equation}
  \label{cH}
  \cH = \frac12\sum_{i<j}\Big[\sin^{-2}\vt_{ij}^-\,(1+\,S_{ij})+\sin^{-2} \vt_{ij}^+\,(1+\tS_{ij})\Big],
\end{equation}
where the lattice sites $\vt_i$ are the coordinates of the unique minimum $\bt$ in the set $A$ of
the scalar potential
\begin{equation}
  \label{UD}
  U(\bx)=\sum_{i\neq
    j}\big(\sin^{-2}x_{ij}^-+\sin^{-2}x_{ij}^+\big).
\end{equation}
Heuristically, the relation between the spin dynamical model~\eqref{H} and its associated spin
chain~\eqref{cH} can be explained as follows. Defining the coordinate-dependent matrix
multiplication operator
\[
h(\bx)=\frac12\sum_{i<j}\big[\sin^{-2} x_{ij}^-\,(1+\,S_{ij})+\sin^{-2}
x_{ij}^+\,(1+\tS_{ij})\big]\,,
\]
the spin Hamiltonian~\eqref{H} can be decomposed as
\begin{equation}
  \label{H3D}
  H = \Hsc+4a\,h(\bx)\,,
\end{equation}
where
\begin{equation}
  \label{Hsc}
  \Hsc=-\sum_i \pa_{x_i}^2 + a(a-1)U(\bx)
\end{equation}
is the Hamiltonian of the scalar Sutherland model of $D_N$ type. Thus, for sufficiently large $a$
all the eigenfunctions of $\Hsc$ are sharply peaked around the unique minimum $\bt$ of the scalar
potential $U$ in the set $A$~\cite{Si83}. Hence, if $\vp_i(\bx)$ is an eigenfunction of $\Hsc$
with energy $\Esc_i$ and $\ket{\si_j}$ is an eigenstate of the chain $\cH$ with eigenvalue
$\cE_j$, for $a\gg 1$ we have
\[
h(\bx)\vp_i(\bx)\ket{\si_j}\simeq\vp_i(\bx)h(\bt)\ket{\si_j}\equiv\vp_i(\bx)\cH\ket{\si_j}
=\cE_j\vp_i(\bx)\ket{\si_j}\,.
\]
By Eq.~\eqref{H3D}, $H$ is approximately diagonal in the basis with elements
$\vp_i(\bx)\ket{\si_j}$, and its eigenvalues $E_{ij}$ satisfy
\begin{equation}
  E_{ij}\simeq \Esc_i+4a\cE_j\,,\qquad a\gg1\,.
  \label{frtr}
\end{equation}
It was shown in Ref.~\cite{CS02} that the scalar potential $U(\bx)$ has a unique minimum in the
configuration space $A$, which coincides with the unique maximum in this set of the ground state
wave function of the scalar Sutherland Hamiltonian of $D_N$ type~\eqref{Hsc}, given by
\begin{equation}
  \label{varrho}
  \rho(\bx) = \prod_{i<j}\big|\sin x_{ij}^-\,\sin x_{ij}^+\big|^a\,.
\end{equation}
The lattice sites $\vt_i$ of the chain~\eqref{cH} are thus the unique solution in $A$ of the
nonlinear system
\begin{equation}
  \label{sitesD}
  \sum_{j;j\ne i}\big(\cot \vt_{ij}^-+\cot \vt_{ij}^+\big)=0\,,\qquad 1\le i\le N\,.
\end{equation}
As in the $BC_N$ case, we define the variables $\xi_i$ by
\[
\xi_i = \cos(2\vt_i)\,,\qquad 1\le i\le N\,.
\]
Note that, since $\bt\in A$, we obviously have
\begin{equation}
  \label{zineq}
  1\ge\xi_1>\xi_2>\cdots >\xi_{N-1}>\xi_N\ge-1\,.
\end{equation}
In terms of the new coordinates $\xi_i$, the system~\eqref{sitesD} can be written as
\begin{equation}
  \label{sitesz}
  (1-\xi_i^2)\sum_{j;j\ne i}\frac1{\xi_i-\xi_j}=0\,,\qquad 1\le i\le N\,.
\end{equation}
Since $\xi_{1}-\xi_j>0$ for all $j>1$ and $\xi_{N}-\xi_j<0$ for all $j<N$, from Eq.~\eqref{sitesz}
it immediately follows that $\xi_{1}^2=\xi_{N}^2=1$, so that $\xi_{1}=-\xi_{N}=1$ by
Eq.~\eqref{zineq}. Substituting into~\eqref{sitesz} we obtain the following system for the
remaining coordinates $\xi_2,\dots,\xi_{N-1}$:
\begin{equation}
  \label{siteszsimp}
  (1-\xi_i^2)\sum_{\substack{j=2\\j\ne i}}^{N-1}\frac1{\xi_i-\xi_j}=2\xi_i\,,\qquad 2\le i\le N-1\,.
\end{equation}
Note that the latter system is invariant under the transformation $\xi_i\mapsto-\xi_i$, so that
(by uniqueness) $\xi_i=\xi_{N+1-i}$. Comparing~\eqref{siteszsimp} with the system
\begin{equation}
  \label{Jacobi}
  2(1-\ze_i^2)\sum_{\substack{j=1\\j\ne
      i}}^{N'}\frac1{\ze_i-\ze_j}=\be-\be'+(\be+\be')\ze_i\,,
  \quad 1\le i\le N'\,,
\end{equation}
satisfied by the zeros $\ze_i$ ($i=1,\dots,N'$) of the Jacobi polynomial $P^{(\be-1,\be'-1)}_{N'}$
(cf.~Ref.~\cite{ABCOP79}), we conclude that the coordinates $\xi_2,\dots,\xi_{N-1}$ are the zeros
of $P_{N-2}^{(1,1)}$. (Note that $P_{N-2}^{(1,1)}$ is proportional to the Gegenbauer polynomial
$C_{N-2}^{(3/2)}$, cf.~Ref.~\cite{Sz75}.) In terms of the original site coordinates $\vt_i$ we
have
\[
0=\vt_1<\vt_2<\dots<\vt_{N-1}<\vt_N=\frac\pi2\,,
\]
with $P_{N-2}^{(1,1)}\big(\cos(2\vt_i)\big)=0$ for $i=2,\dots,N-1$. Note that
Eqs.~\eqref{sitesz}-\eqref{Jacobi} also yield an alternative characterization of the coordinates
$\xi_i$ as the $N$ roots of the Jacobi polynomial $P_N^{(-1,-1)}$, which was to be expected, since
the potential $U^{(\mathrm B)}$ in Eq.~\eqref{UB} reduces to the $D_N$ potential $U$ when
$\be=\be'=0$. The equivalence of both characterizations of the site coordinates is easily
established with the help of the identity $4P_N^{(-1,-1)}(t)=(t^2-1)P_{N-2}^{(1,1)}(t)$,
cf.~\cite{OS02}.

We shall next discuss the precise relation between the $D_N$ spin chain Hamiltonian~\eqref{cH} and
the limit as $(\be,\be')\to0$ of its $BC_N$ counterpart~\eqref{HSchain}. To this end, we use the
trigonometric identities
\[
\sin^{-2}\te_i=\frac2{1-\ze_i}\,,\qquad \cos^{-2}\te_i=\frac2{1+\ze_i}\,,
\]
and note that as $(\be,\be')\to0$ all the roots $\ze_i$ of the Jacobi polynomial
$P_N^{(\be-1,\be'-1)}$ tend to the corresponding roots $\xi_i$ of $P_N^{(-1,-1)}$. Thus all the
terms in the last sum in the Hamiltonian~\eqref{HSchain} tend to zero as $(\be,\be')\to(0,0)$,
except the first and the last one. In order to evaluate the limit of these two terms, we
divide~\eqref{Jacobi} by $1\pm \ze_i$ and sum the resulting equation over $i$, obtaining
\[
2\be\sum_i\frac1{1-\ze_i}=2\be'\sum_i\frac1{1+\ze_i}=N(\be+\be'+N-1)\,.
\]
Hence
\begin{equation}\label{limits}
  \lim_{(\be,\be')\to0}\frac{2\be}{1-\ze_1}=\lim_{(\be,\be')\to0}\frac{2\be'}{1+\ze_
    N}=N(N-1)\,.
\end{equation}
Since $\ze_i\to \xi_i$ as $(\be,\be')\to0$ and $\te_i,\vt_i\in A$ (recall that $A^{(B)}\subset
A$), we have $\lim_{(\be,\be')\to0}\te_i=\vt_i$ for all $i=1,\dots,N$. {}From
Eqs.~\eqref{HSchain}, \eqref{cH} and~\eqref{limits} it immediately follows that
\begin{equation}\label{impurity}
  \lim_{(\be,\be')\to0}\cH^{(\mathrm B)}=\cH+\frac12\,N(N-1)\Big[1-\frac\ep2(S_1+S_N)\Big].
\end{equation}
Thus the limit as $(\be,\be')\to0$ of the Hamiltonian of the HS chain of $BC_N$ type yields its
$D_N$ analog, plus an additional term which can be interpreted as an ``impurity'' at both ends of
the latter chain.

\section{Spectrum of the dynamical models}\label{sec.spectrum}
The aim of this section is to compute the spectra of the su($m$) spin Sutherland model of $D_N$
type~\eqref{H} and its scalar counterpart~\eqref{Hsc}. In order to facilitate this computation, we
introduce the auxiliary \emph{scalar} operator
\begin{equation}
  \label{Hp}
  H' = -\sum_i \pa_{x_i}^2 + a\,\sum_{i\neq j}\big[\sin^{-2}x_{ij}^-\,(a-K_{ij})
  +\sin^{-2} x_{ij}^+\,(a-\tK_{ij})\big]\,,
\end{equation}
where $K_{ij}$ and $K_i$ are coordinate permutation and sign reversing operators, defined by
\begin{align*}
  &(K_{ij}f)(x_1,\dots,x_i,\dots,x_j,\dots,x_N)=f(x_1,\dots,x_j,\dots,x_i,\dots,x_N)\,,\\
  &(K_i f)(x_1,\dots,x_i,\dots,x_N)=f(x_1,\dots,-x_i,\dots,x_N)\,,
\end{align*}
and $\tK_{ij}\equiv K_iK_jK_{ij}$.

\subsection{Extensions of $H$ and $\Hsc$}\label{subsec.ext}

Due to the character of their singularities, the operators $H$ and $\Hsc$ are naturally defined on
suitable dense subspaces of the Hilbert spaces $L^2(A)\otimes\Si$ and $L^2(A)$, respectively. On
the other hand, the appearance in the RHS of Eq.~\eqref{Hp} of the generators $K_iK_j$ and
$K_{ij}$ of the Weyl group $\fW$ of the $D_N$ root system entails that the auxiliary operator $H'$
is defined instead on a dense subspace of $L^2(C)$, where $C\equiv\fW(A)$. One of the key
ingredients of the method we shall use consists in replacing the operators $H$ and $\Hsc$ by
suitable equivalent (isospectral) extensions $\tH$ and $\tHsc$ thereof to appropriate subspaces of
$L^2(C)\otimes\Si$ and $L^2(C)$, such that $\tH=H'\otimes\id$ and $\tHsc=H'$ in the latter
subspaces.

We shall start by showing that the set $C$ is explicitly given by
\begin{equation}
  \label{Ceq}
  C=\big\{\bx\in\RR^N:0<|x_i\pm x_j|<\pi\,,\enspace 1\le i<j\le N\big\}\,,
\end{equation}
a characterization which will prove useful in what follows. Recall, to this end, that $\fW$ is
generated by coordinate permutations and sign reversals of an \emph{even} number of
coordinates~\cite{Hu90}. Since $A$ is defined by the inequalities~\eqref{defA}, it is obvious that
$C\subset C^*$, where $C^*$ denotes the RHS of Eq.~\eqref{Ceq}. Thus, we need only show that
$C^*\subset C$, or equivalently, that for every $\bx^*\in C^*$ there is an element $W\in\fW$ such
that $W\bx^*\in A$. To this end, we shall make repeated use of the following elementary fact:
\begin{equation}\label{obs}
  \bx\in C^*,\enspace |x_1|<x_2<\cdots<x_N\quad\implies\quad\bx\in A.
\end{equation} 
Suppose, then, that $\bx^*\in C^*$. By reversing the sign of an even number of coordinates and
applying a suitable permutation, we can transform $\bx^*$ into another element
$\by\in\fW(C^*)=C^*$ satisfying
\[
y_1<y_2<\dots<y_N\quad\text{and}\quad y_i>0\,,\qquad i=2,\dots,N\,.
\]
If $y_1+y_2>0$, then $|y_1|<y_2<\dots<y_N$, and hence $\by\in A$ by Eq.~\eqref{obs}. Suppose, on
the other hand, that $y_1+y_2<0$, so that
\[
y_1<0<y_2<\dots<y_N\,.
\]
Calling $z_1=-y_2$, $z_2=-y_1$ and $z_i=y_i$ for $i\ge 3$, we have $\bz\in C^*$ and
\[
z_1+z_2=-(y_1+y_2)>0\,,\quad z_1<z_2\quad\implies\quad |z_1|<z_2.
\]
If $z_2<z_3$, then $\bz\in A$ by Eq.~\eqref{obs}. Otherwise, the inequalities
$z_1+z_3=y_3-y_2>\nobreak 0$ and $z_3-z_1=y_3+y_2>0$ imply that
\[
|z_1|<z_3<\cdots<z_2<\cdots<z_N\,.
\]
Applying a suitable permutation to $\bz$, we obtain a new element $\bu\in C^*$ satisfying $
|u_1|<u_2<\cdots<u_N $, which belongs to $A$ again by Eq.~\eqref{obs}.
\begin{remark}
  Note that the analogous set $C^{(\mathrm B)}$ for the $BC_N$ root system, which is simply the
  hypercube $(-\frac\pi2,\frac\pi2)^N$ minus the singular hyperplanes $x_i\pm x_j=0$, $1\le i\le
  j\le N$ (cf.~Ref.~\cite{EFGR05}), is clearly contained in $C$ by Eq.~\eqref{Ceq}.
\end{remark}
\begin{remark}
  For $N=3$, the set $\BC$ is a \emph{rhombic dodecahedron} (a zonohedron with 12 equal rhombic
  faces) centered at the origin~\cite{Co73}, with edge length $\sqrt 3\ms\pi/2$.
\end{remark}

Our next aim is to replace the operator $H$ by an isospectral extension $\tH$ thereof acting on
suitably (anti)symmetrized wave functions defined on the set~$C$. If the extension is
appropriately chosen, we shall see that $\tH=H'\otimes\id$ on the Hilbert space of $\tH$, so that
the computation of the spectrum of $H$ reduces to the analogous (but considerably simpler, in
practice) task for $H'$.

Before proceeding with our construction, we need to introduce some additional notation. Given an
element $W$ of the Weyl group of $D_N$ type $\fW$ and a factorized spin function $\vp\ket\bs\in
L^2(C)\otimes\Si$, where $\ket\bs\equiv\ket{s_1,\dots,s_N}$ is an element of the canonical spin
basis, we define the action of $W$ on $\vp\ket\bs$ in the usual way:
\begin{equation}
  W(\vp\ket\bs)=(\vp\circ W^{-1})\ket{W\bs}\,.
  \label{Wphkets}
\end{equation}
Extending this definition by linearity to the whole Hilbert space $L^2(C)\otimes\Si$, we obtain an
action of $\fW$ (in fact, a representation) on the latter space. Let now $\pi:\fW\to\CC$ denote a
one-dimensional representation of $\fW$; of course, since $\fW$ is generated by reflections, we
have $\pi(\fW)\subset\{-1,1\}$. The \emph{symmetrizer} $\La_\pi$ associated with $\pi$ is the
linear operator defined on $L^2(C)\otimes\Si$ by
\begin{equation}
  \La_\pi=\frac1{|\fW|}\sum_{W\in\fW}\pi(W)\ms W\,,
  \label{Lapi}
\end{equation}
where $|\fW|=2^{N-1}N!$ is the order of $\fW$. By construction, we have
\begin{equation}
  \Phi\in\La_\pi\big(L^2(C)\otimes\Si\big),\enspace W\in\fW\quad\implies\quad
  W\Phi=\pi(W)\ms\Phi\,,
  \label{proj}
\end{equation}
so that $\La_\pi$ is the projector onto states with well-defined parity $\pi(W)$ with respect to
any transformation $W$ in the Weyl group $\fW$.

Given a factorized spin function $\Psi=\psi\ket\bs\in L^2(A)\otimes\Si$, we define its
\emph{$\pi$-symmetric extension} $\tilde\Psi\in\La_\pi\big(L^2(C)\otimes\Si\big)$ by
\begin{equation}
  \label{ext}
  \tilde\Psi(\bx)=\pi(W_\bx)\,\psi(W^{-1}_\bx\bx)\ms\ket{W_\bx\bs}\,,\qquad \bx\in C\,,
\end{equation}
where $W_\bx$ denotes the \emph{unique} element of $\fW$ such that $W_\bx^{-1}\bx\in A$. As usual,
the action of $\enspace\widetilde{\vphantom a}\enspace$ is extended to $L^2(A)\otimes\Si$ by
linearity. It is easy to see that $\tilde\Psi$ is the unique extension of $\Psi$ to $C$ which has
well defined parity $\pi(W)$ under any transformation $W\in\fW$. In view of
Eqs.~\eqref{Wphkets}-\eqref{Lapi}, with a slight abuse of notation we can write
\begin{equation}
  \tilde\Psi=|\fW|\cdot\La_\pi(\Psi\chi_A)\,,
  \label{tpsiLa}
\end{equation}
where $\chi_A$ is the characteristic function of $A$.

The extension $\enspace\widetilde{\vphantom
  a}\enspace:L^2(A)\otimes\Si\to\La_\pi\big(L^2(C)\otimes\Si\big)$ is an invertible linear
operator, its inverse being the restriction operator $\enspace\widehat{\vphantom
  a}\enspace:\La_\pi\big(L^2(C)\otimes\Si\big)\to L^2(A)\otimes\Si$ defined by $\hat\Phi=\Phi|_A$.
Indeed, the linearity of the operator $\enspace\widetilde{\vphantom a}\enspace$ is obvious. As to
its invertibility, note first of all that if $\Psi\in L^2(A)\otimes\Si$ by Eq.~\eqref{ext} we have
$\hat{\tilde\Psi}=\Psi$, since $W_\bx$ is the identity when $\bx$ belongs to $A$. On the other
hand, if $\Phi\in\La_\pi\big(L^2(C)\otimes\Si\big)$ using Eqs.~\eqref{Lapi} and~\eqref{tpsiLa} we
obtain
\[
  \tilde{\hat\Phi}=\sum_{W\in\fW}\pi(W)\ms W(\hat\Phi\chi_A)=\sum_{W\in\fW}\pi(W)\ms W(\Phi\chi_A)
  =\sum_{W\in\fW}\pi(W)\ms W(\Phi)\,\Big(\chi_A\circ W^{-1}\Big)\,,
\]
and hence, by Eq.~\eqref{proj},
\[
\tilde{\hat\Phi}(\bx)=\Phi(\bx)\sum_{W\in\fW}\chi_A(W^{-1}\bx)=\Phi(\bx)\chi_A(W_\bx^{-1}\bx)
=\Phi(\bx)\,.
\]

Given a one-dimensional representation $\pi$ of $\fW$ and a linear operator $T$ acting on
$L^2(A)\otimes\Si$, it is natural to define its \emph{$\pi$-symmetric extension} $\tilde T_\pi$ to
the Hilbert space $\La_\pi\big(L^2(C)\otimes\Si\big)$ by the prescription
\begin{equation}
  \label{Hext}
  \tilde T_\pi\,\Phi = (T\ms\hat\Phi)^{\widetilde{\phantom a}}\,,
  \qquad \Phi\in \La_\pi\big(L^2(C)\otimes\Si\big)\,.
\end{equation}
By the invertibility of the $\enspace\widetilde{}\enspace$ operator, we have
\[
\tilde T_\pi =\widetilde{\phantom a}\circ T\circ(\widetilde{\phantom a})^{-1}\,,
\]
so that the operators $T$ and $\tilde T_\pi$ are isospectral.

We now seek to find a suitable one-dimensional representation $\pi$ of $\fW$ such that the
$\pi$-symmetric extension of $H$ coincides with the restriction of $H'\otimes\id$ to
$\La_\pi\big(L^2(C)\otimes\Si\big)$. In view of Eqs.~\eqref{H}-\eqref{Hp}, it suffices that $\pi$
satisfy
\[
K_iK_j\La_\pi=S_iS_j\La_\pi\,,\qquad K_{ij}\La_\pi=-S_{ij}\La_\pi\,.
\]
Hence $\pi(W)$ should be defined as the sign of the permutation part of $W\in\fW$. {}From now on,
when dealing with the spin Hamiltonian~\eqref{H} we shall take $\pi$ as above, and drop the
subscript $\pi$ from $\La_\pi$ and~$\tH_\pi$.

Turning now to the scalar Hamiltonian $\Hsc$, by Eqs.~\eqref{UD}, \eqref{Hsc}, and \eqref{Hp} its
extension $\tHsc$ to the space $\La_\pi\big(L^2(C)\big)$ will coincide with the restriction of
$H'$ to the latter space provided that
\[
K_iK_j\La_\pi=\La_\pi\,,\qquad K_{ij}\La_\pi=\La_\pi\,.
\]
Thus in this case $\pi(W)=1$ for all $W\in\fW$, so that for the scalar model~\eqref{Hsc}
$\La_\pi\equiv\Lsc$ is the total symmetrizer with respect to both coordinate permutations and sign
reversals of an even number of coordinates.

By the previous discussion, we have reduced the problem of evaluating the spectrum of the
Hamiltonians $\Hsc$ and $H$ to the analogous problem for the restrictions of the auxiliary
operators $H'$ and $H'\otimes\id$ to the Hilbert spaces $\Lsc\big(L^2(C)\big)$ and
$\La\big(L^2(C)\otimes\Si\big)$, respectively. We shall next prove a more explicit
characterization of these spaces that will be needed in the sequel. To this end, let $\La^\pm$ be
the projector onto states antisymmetric under particle permutations and with parity $\pm1$ under
reversals of coordinates and spins. We shall show that
\begin{equation}
  \La\big(L^2(C)\otimes\Si\big)=\La^+\big(L^2(C)\otimes\Si\big)\oplus
  \La^-\big(L^2(C)\otimes\Si\big)\,.
  \label{LaL2Si}
\end{equation}
Indeed, let $\La_{\mathrm a}$ denote the antisymmetrizer under particle permutations, and let
$\big\{W_i^{\mathrm \pm}\big\}_{i=1}^{2^{N-1}}$ be the set of reversals of an even ($+$) or an odd
($-$) number of coordinates and spins. We then have
\[
\La^\pm=\frac1{2^N}\bigg(\sum_{i=1}^{2^{N-1}}W^+_i\pm\sum_{i=1}^{2^{N-1}}W^-_i\bigg)\La_{\mathrm a}\,,
\]
and hence
\[
\La=\frac1{2^{N-1}}\bigg(\sum_{i=1}^{2^{N-1}}W^+_i\bigg)\La_{\mathrm a}=\La^++\La^-\,,
\]
which establishes~\eqref{LaL2Si}. Similarly, if $\Lsc^\pm$ is the projector onto states symmetric
under coordinate permutations and with parity $\pm1$ under sign reversals, it is easy to show that
\begin{equation}
  \Lsc\big(L^2(C)\big)=\Lsc^+\big(L^2(C)\big)\oplus \Lsc^-\big(L^2(C)\big)\,.
  \label{LaL2}
\end{equation}

\subsection{Basis of $L^2(C)$}

Our next step is to construct suitable (non-orthogonal) bases\footnote{%
  More precisely, a non-orthogonal basis of a (separable) Hilbert space $\fH$ is a Schauder basis
  of its underlying Banach space, i.e., a countable subset $\{v_i\;:\; i\in\NN\}\subset\fH$ such
  that every element $v\in\fH$ can be expressed in a unique way as $\sum_{i=1}^\infty c_i\ms v_i$,
  with $c_i\in\CC$. In the rest of this section the term ``basis'' will often be used in this more
  general sense.}%
of $\La\big(L^2(C)\otimes\Si\big)$ and $\Lsc\big(L^2(C)\big)$ on which $H'\otimes\id$ and $H'$,
respectively, act as triangular matrices. The decompositions~\eqref{LaL2Si}-\eqref{LaL2}, and the
fact that $H'\otimes\id$ and $H'$ clearly commute with $\La^\pm$ and $\Lsc^\pm$, suggest that we first
triangularize $H'$ on $L^2(C)$. It should be noted that this problem is considerably harder than
the corresponding one for the rational $D_N$ model studied in Ref.~\cite{BFG09}, due to the fact
that in the latter case $\BC=\BC^{(\mathrm{B})}=\RR^N$. However, in the present case $\BC$ does
not coincide with $\BC^{(\mathrm{B})}=[-\frac\pi2,\frac\pi2]^N$, so that one cannot assume that
the functions
\begin{equation}\label{BCbasis}
  \rho(\bx)\,\e^{2\iu\ms\bn\cdot\bx}\,,\qquad \bn\in\ZZ^N\,,
\end{equation}
obtained from the basis of $L^2\big([-\frac\pi2,\frac\pi2]^N\big)$ found in Ref.~\cite{EFGR05} by
setting $b=b'=0$, are a basis of $L^2(C)$. In fact, it turns out that the set~\eqref{BCbasis} is
\emph{not} complete in $L^2(C)$, and must therefore be supplemented by additional functions in
order to obtain a basis. This peculiarity, which is absent in the rational case, lends an
additional layer of complexity to the relation between the trigonometric $D_N$ models and their
$BC_N$ counterparts.

In this subsection we shall prove that the functions
\begin{equation}
  \label{Dbasis}
  \vp_{\bn}^{(\de)}(\bx)\equiv\rho(\bx)\,\e^{\iu\ms\sum_j(2n_j+\de) x_j}\,,
  \qquad\bn\equiv(n_1,\dots,n_N)\in\ZZ^N\,,\quad \de\in\{0,1\}\,,
\end{equation}
form a Schauder basis of $L^2(C)$, leaving for the next subsection the proof that $H'$ acts
triangularly on the latter set when it is ordered appropriately. As we shall next discuss in more
detail, the completeness of the functions~\eqref{Dbasis} is essentially based on the fact that a
complex exponential $\e^{\iu\ms\bk\cdot\bx}$ ($\bk\in\RR$) is periodic in $\BC$ if and only if
\begin{equation}\label{ks}
  \bk=(2n_1+\de,\dots,2n_N+\de),\qquad n_j\in\ZZ,\quad \de\in\{0,1\}\,.
\end{equation}

Since $\BC$ is not a hypercube, we need to define more precisely what it means for a function to
be periodic in this set. To this end, let
\[
F_{ij}^{\ep\ep'}=\big\{\bx\in\BC:x_i+\ep\ms x_j=\ep'\pi\big\}\,,\qquad 1\le i<j\le
N\,,\quad\ep,\ep'=\pm\,,
\]
denote one of the $2N(N-1)$ faces of $\BC$. If $T_{ij}^{\ep\ep'}$ is the translation along the
vector $\ep'\pi(\bev_i+\ep\ms \bev_j)$ perpendicular to the latter face (where
$\{\bev_1,\dots,\bev_N\}$ is the canonical basis of $\RR^N$), then each $T_{ij}^{\ep,-\ep'}$
clearly sends the corresponding face $F_{ij}^{\ep\ep'}$ to its opposite $F_{ij}^{\ep,-\ep'}$.
Given a point $\bx\in F_{ij}^{\ep\ep'}$, we shall refer to $T_{ij}^{\ep,-\ep'}\bx$ as the point
opposite to $\bx$ in the face $F_{ij}^{\ep,-\ep'}$. (Of course, a point lying on the intersection
of $k>1$ faces has $k$ different opposites.)

We shall say that a continuous function $f:\BC\to\CC$ is \emph{periodic} in $\BC$ if
\begin{equation}
  f(\bx)=f\big(T_{ij}^{\ep,-\ep'}\bx\big)\,, \qquad\forall\bx\in F_{ij}^{\ep\ep'}\,,\quad 1\le
  i<j\le N\,,\quad\ep,\ep'=\pm\,.
  \label{Cper}
\end{equation}
In other words, $f$ is periodic in $\BC$ if it takes the same value on opposite points in any two
faces of $\BC$. Since the coroots of the $D_N$ root system with the normalization~\eqref{DNroots}
are the $2N(N-1)$ vectors
\[
\pi(\pm \bev_i\pm \bev_j)\,,\qquad 1\le i<j\le N\,,
\]
the group $\fT$ generated\footnote{In fact, since $T_{kj}^{+,\mp}T_{ik}^{+,\pm}=T_{ij}^{-,\pm}$ for
  $i<k<j$, the group $\fT$ is generated just by the translations $T_{ij}^{+,\pm}$.} by the
translations $T_{ij}^{\ep\ep'}$ is the translation group corresponding to the $D_N$ coroot lattice
(i.e., the $\ZZ$-linear span of the coroot vectors). As is well known, the semidirect product of
$\fT\ltimes\fW$ yields the affine Weyl group of $D_N$ type $\fWa$~\cite{Hu90}. We shall define two
points $\bx,\bx'\in\RR^N$ to be \emph{equivalent}, and shall write $\bx\sim\bx'$, provided that
$\bx'=T\bx$ for some $T\in\fT$. Note that Eq.~\eqref{Cper} and the previous definition imply that
a function $f$ is periodic in $\BC$ if and only if
\[
\bx,\bx'\in\pa\BC\,,\; \bx\sim\bx'\implies f(\bx)=f(\bx')\,.
\]
We shall accordingly say that a continuous function $f:\RR^N\to\CC$ is $\fT$-periodic if it
satisfies
\begin{equation}\label{fper}
  f(\bx)=f(T\bx)\,,\qquad\forall T\in\fT\,,\;\forall\bx\in\RR^N\,.
\end{equation}
\begin{remark}
  \label{rem.2piper}
  Since $T_{ij}^{++}T_{ij}^{-+}$ is a translation of $2\pi$ in the direction of the vector
  $\bev_i$, it follows that every $\fT$-periodic function is $2\pi$-periodic in each coordinate.
  The converse, however, is not true in general.
\end{remark}
One of the main ingredients in the proof of the completeness of the set~\eqref{Dbasis} in $L^2(C)$
is the fact that every continuous function $f:\BC\to\CC$ periodic in $\BC$ can be uniquely
extended to a $\fT$-periodic function $\bar f:\RR^N\to\CC$. This result is a direct consequence of
the following fundamental facts:
  \begin{enumerate}[i)]
  \item For each $\bx\in\RR^N$, there is a point $\bx'\in\BC$ such that
    $\bx\sim\bx'$.\smallskip
  \item  Moreover, if $\bx\sim\bx''\in\BC$ and $\bx''\ne\bx'$, then both $\bx'$ and
    $\bx''$ lie on \textup(at least\textup) a face of $\BC$.
  \end{enumerate}
  The proof of these two statements is straightforward. Indeed, it is well known~\cite{Hu90} that
  $\BA$ is a fundamental domain for the action of $\fWa$ in $\RR^N$, i.e., that for every
  $\bx\in\RR^N$ there is a unique $\ba\in\BA$ and a suitable element $R$ of $\fWa$ such that
  $\bx=R\ba$. Since $\fWa$ is the semidirect product of its subgroups $\fT$ and $\fW$, we can
  write $R=TW$, with $T\in\fT$ and $W\in\fW$. This shows that $\bx$ is equivalent to
  $\bx'=W\ba\in\fW(\BA)$. Since the elements of $\fW$ are homeomorphisms, we have
  $\fW(\BA)=\overline{\fW(A)}\equiv\BC$, which proves the first statement.

As to the second one, suppose next that $\bx$ is equivalent to two different points $\bx'$ and
$\bx''$ of $\BC$. It follows that $\bx'\sim\bx''$ or, equivalently,
\[
x_i''=x_i'+k_i\pi\,,\qquad k_i\in\ZZ\,,\ 1\le i\le N\,;\quad k_1+\cdots+k_N\in2\ZZ\,.
\]
Since both $\bx'$ and $\bx''$ belong to $\BC\subset[-\pi,\pi]^N$, the integers $k_i$ can only take
the values $0,\pm1,\pm2$. Suppose, first, that one of these integers is equal to $\pm2$. Without loss of
generality, we may assume that $k_1=2$. Since $x_1',x_1''\in[-\pi,\pi]$, this is only possible if
$x_1'=-\pi=-x_1''$, from which it follows (taking into account that $\bx',\bx''\in\BC$) that
$x_i'=x_i''=0$ for $i>1$. Thus in this case $\bx'$ belongs to the $2(N-1)$ faces $F_{1i}^{\pm,-}$
(with $2\le i\le N$), and $\bx''$ to the opposite faces $F_{1i}^{\pm,+}$. (Note, however, that in
this case $\bx''$ is \emph{not} the opposite point of $\bx'$ in any of these $2(N-1)$ faces.)

Assume now that $|k_i|<2$ for all $i$. Since not all the integers $k_i$ can be zero by hypothesis,
and $k_1+\cdots +k_N$ must be even, there are two of these integers, which
w.l.o.g.~we can take as $k_1$ and $k_2$, such that $k_1=\ep k_2=\ep'$ (with $\ep,\ep'=\pm$). From
the equalities $x_1''+\ep\ms x''_2=x_1'+\ep\ms x_2'+2\ep'\pi$ and the fact that $x_1'+\ep
x_2',x_1''+\ep\ms x_2''\in[-\pi,\pi]$, it follows that $x_1''+\ep\ms x_2''=-(x_1'+\ep
x_2')=\ep'\pi$, and therefore $\bx'$ and $\bx''$ lie on (at least) the opposite faces
$F_{12}^{\ep\ep'}$ and $F_{12}^{\ep,-\ep'}$ of $\BC$, respectively. This completes the proof of
the second statement.
\begin{remark}\label{rem.tiling}
  The previous result implies that $N$-dimensional Euclidean space can be tiled with
  translations of the set $\BC$ along the $D_N$ (co)root lattice. For $N=3$, this is the
  well-known tessellation of $\RR^3$ with rhombic dodecahedra~\cite{Co73}.
\end{remark}
We are now ready to prove that the set~\eqref{Dbasis} is a (non-orthogonal) basis of $L^2(C)$. In
fact, it suffices to show that the exponentials
\begin{equation}
  \label{fourierD}
  \e^{\iu\ms\sum_j(2n_j+\de) x_j}\,,
  \qquad\bn\equiv(n_1,\dots,n_N)\in\ZZ\,,\quad \de\in\{0,1\}
\end{equation}
are themselves a basis of $L^2(C)$. Indeed, if this is the case then any complex-valued function
$f\in C_0(\BC)$ continuous in $\BC$ and with compact support in $C$ can be represented in the form
\begin{equation}
  \label{fseries}
  f(\bx)=\sum_{\bn\in\ZZ^N,\,\de\in\{0,1\}}c_{\bn,\de}\ms\e^{\iu\ms\sum_j(2n_j+\de) x_j}\,,
\end{equation}
where the coefficients $c_{\bn,\de}\in\CC$ are by hypothesis uniquely determined by $f$. Our claim
follows immediately from the fact that the function $f/\rho$ is also in $C_0(\BC)$, and that the
latter set is of course dense in $L^2(\BC)$.

The fact the exponentials~\eqref{fourierD} form a basis of $L^2(C)$ is essentially a consequence
of the fact that the momenta~\eqref{ks} are the elements of the dual (or ``reciprocal'', in a more
physical terminology) lattice of the $D_N$ coroot lattice. However, for completeness' sake we
shall next provide an elementary proof of this fact. We first note that, since the set $P(\BC)$ of
complex-valued continuous functions periodic in $\BC$ contains the dense set $ C_0(\BC)$, to prove
that~\eqref{fourierD} is a basis of $L^2(\BC)$ we need only show that every $f\in P(\BC)$ can be
uniquely represented by a Fourier series of the form~\eqref{fseries}. Let, then, $f:\BC\to\CC$ be
a continuous function periodic in $\BC$, and denote by $\bar f:\RR^N\to\CC$ its $\fT$-periodic
extension. Since the function $\bar f$ is $2\pi$-periodic in each coordinate
(cf.~Remark~\ref{rem.2piper}), it can be developed in terms of the $L^2([-\pi,\pi]^N$) Fourier
basis $\e^{\iu\ms\bk\cdot\bx}$ ($\bk\in\ZZ^N$) as
\begin{equation}
  \bar f(\bx)=\sum_{\bk\in\ZZ^N}a_{\bk}\ms\e^{\iu\bk\cdot\bx}\,,
  \label{barfser}  
\end{equation}
with $a_{\bk}\in\CC$ uniquely determined by $f$. Imposing that $\bar f$ satisfy Eq.~\eqref{fper}
when $T$ is one of the generators $T_{ij}^{\ep\ep'}$ of $\fT$ we easily obtain
\[
a_{\bk}\big(1-\e^{\iu\ms\pi\ep'(k_i+\ep k_j)}\big)=0\,, \qquad \forall\bk\in\ZZ^N\,,\enspace 1\le
i<j\le N\,,\enspace \ep,\ep'=\pm\,.
\]
Hence $a_\bk$ vanishes unless
\[
k_i\pm k_j\equiv k_{ij}^\pm\in2\ZZ\,,\qquad 1\le i<j\le N\,,
\]
i.e., unless all the integers $k_i$ have the same parity. Setting $k_i=2n_i+\de$ (with
$\de\in\{0,1\}$), $a_{\bk}=c_{\bn,\de}$, and substituting into Eq.~\eqref{barfser} we obtain
\[
\bar f(\bx)=\sum_{\bn\in\ZZ^N,\,\de\in\{0,1\}}c_{\bn,\de}\ms\e^{\iu\ms\sum_j(2n_j+\de) x_j}\,,
\]
where the equality should be understood in the sense of $L^2([-\pi,\pi]^N)$. Restricting to
$L^2(\BC)$ (which is allowed, since $\BC\subset[-\pi,\pi]^N$), and recalling that all the
coefficients $a_{\bk}$ are uniquely determined by $f$, we obtain the desired result.
\begin{remark}
  In fact, since the translations of $\BC$ along the $D_N$ (co)root lattice are a tiling of $R^N$,
  the results of~\cite{Fu74} imply that the functions~\eqref{fourierD} are mutually orthogonal. Of
  course, this does not imply that the functions~\eqref{Dbasis} are themselves orthogonal, due to
  the presence of the factor $\rho(\bx)$.
\end{remark}

\subsection{Triangularization of $H'$}
We shall next endow the set~\eqref{Dbasis} with a suitable order such that the action of $H'$ on
the resulting basis is triangular. Note, first of all, that
\begin{equation}\label{V0V1}
  L^2(C)=\fH^{(0)}\oplus\fH^{(1)}\,,
\end{equation}
where $\fH^{(\de)}$ is the closure of the subspace spanned by the basis functions
$\vp^{(\de)}_\bn$ with $\bn\in\ZZ^N$. We will show that $H'$ leaves invariant each of the
subspaces $\fH^{(\de)}$, so that we need only order each subbasis
$\big\{\vp_\bn^{(\de)}\big\}_{\bn\in\ZZ^N}$ in such a way that $H'$ is represented by a triangular
matrix in $\fH^{(\de)}$. To this end, given a multiindex $\bp\equiv(p_1,\dots,p_N)\in\ZZ^N$ we
define
\[ [\bp] = \big(|p_{i_1}|,\dots,|p_{i_N}|\big)\,,\qquad
\text{with}\quad|p_{i_1}|\ge\cdots\ge|p_{i_N}|\,.
\]
If $\bp'\in\ZZ^N$ is another multiindex, we shall write $\bp\prec\bp'$ provided that the first
non-vanishing component of $[\bp]-[\bp']$ is negative. The basis functions
$\big\{\vp_\bn^{(\de)}\big\}_{\bn\in\ZZ^N}$ should then be ordered in any way such that
$\vp_\bn^{(\de)}$ precedes $\vp_{\bn'}^{(\de)}$ whenever $\bnu\prec \bnu'$, where
\begin{equation}\label{bnu}
  \bnu\equiv (2n_1+\de,\dots,2n_N+\de)\,,
\end{equation}
and similarly for $\bnu'$. For instance, $\vp_{(3,1,0)}^{(0)}$ must precede
$\vp_{(2,-3,-1)}^{(0)}$, $\vp_{(3,1,0)}^{(1)}$ should follow $\vp_{(2,-3,-1)}^{(1)}$, while the
relative precedence of $\vp_{(2,-3,-1)}^{(0)}$ and $\vp_{(1,3,-2)}^{(0)}$ can be arbitrarily
assigned.

In order to compute the action of $H'$ on the basis functions~\eqref{Dbasis}, it is convenient to
introduce the Dunkl operators of $D_N$ type
\begin{equation}\label{J}
  J_k=\iu\,\pa_{x_k}+a\sum_{l\neq k}\Big[(1-\iu\cot
  x_{kl}^-)\,K_{kl}+(1-\iu\cot x_{kl}^+)\,\tK_{kl}\Big]-2a\sum_{l<k}K_{kl}\,,
\end{equation}
with $k=1,\dots,N$, obtained from their $BC_N$ counterparts~\cite{EFGR05} by setting $b=b'=0$.
Note that the set $C$ is invariant under {\em all} the generators $K_{ij}$, $K_i$ of the Weyl
group of $BC_N$ type $\fW^{(\mathrm B)}$, and hence the operator $J_k$ is actually defined in a
suitable dense subspace of $L^2(C)$. It can be shown that
\begin{equation}\label{HpJs}
  H'=\sum_k J_k^2\,,
\end{equation}
so that the action of $H'$ on the basis~\eqref{Dbasis} can be easily inferred from that of the
Dunkl operators~\eqref{J}. In the following discussion, we shall label the basis functions
$\vp_{\bn}^{(\de)}$ simply by $\vp_{\bnu}$, with $\bnu$ given by~\eqref{bnu}. As in
Ref.~\cite{FGGRZ03}, we shall begin by considering the action of $J_k$ on a basis functions
$\vp_{\bnu}$ with $\bnu$ nonnegative and nonincreasing. For such a multiindex, we shall use the
notation
\[
\#(s) = \card\{i:\nu_i=s\}\,,\qquad \ell(s) = \min\{i:\nu_i=s\}\,,
\]
with $\ell(s)=+\infty$ if $\nu_i\ne s$ for all $i=1,\dots,N$. For instance, if
$\bnu=(8,6,6,2,2,2)$ then $\#(2)=3$ and $\ell(2)=4$.

We shall next prove the formula
\begin{equation}
  \label{Jifnstruct}
  J_k \vp_{\bnu} = \la_{\bnu,k}\,
  \vp_{\bnu} + \sum_{\substack{\bnu'\in\ZZ^N\\ \bnu'-\bnu\in(2\ZZ)^N,\,\bnu'\!\prec\,\bnu}}
  c_{\bnu,k}^{\bnu'}\,\vp_{\bnu'}\,,
\end{equation}
where $c_{\bnu,k}^{\bnu'}\in\CC$ and
\begin{equation}\label{la}
  \la_{\bnu,k} =
  \begin{cases}
    -\nu_k+2a\big(2\ell(\nu_k)+\#(\nu_k)-k-N-1\big),\quad& \nu_k>0\\[6pt]
    2a(N-k)\,,\quad& \nu_k=0\,,
  \end{cases}
\end{equation}
which will play a fundamental role in the sequel. To begin with, a lengthy but straightforward
calculation yields
\begin{multline}\label{Jkphinu}
  \frac{J_k\vp_{\bnu}}{\vp_{\bnu}} =-\nu_k-2a(N-1)+2a\sum_{j<k}
  \frac{\al_{jk}^{\nu_j-\nu_k}-1}{\al_{jk}^2-1}\\
  +2a\sum_{j>k}\frac{\al_{jk}^{\nu_j-\nu_k+2}-1}{\al_{jk}^2-1} +2a\sum_{j\ne k}
  \frac{\be_{jk}^{2-\nu_j-\nu_k}-1}{\be^2_{jk}-1}\,,
\end{multline}
where
\[
\al_{jk}=z_j^{-1}z_k\,,\qquad \be_{jk}=z_jz_k\,,\qquad z_j\equiv\e^{\iu x_j}\,.
\]
Consider now the first sum in Eq.~\eqref{Jkphinu}. Since $j<k$, by hypothesis $\nu_j\geq\nu_k$.
If $\nu_j=\nu_k$, the $j$-th term in this sum clearly vanishes. On the other hand, if
$\nu_j>\nu_k$ we have
\begin{equation}\label{firstterm}
  \vp_{\bnu}\frac{\al_{jk}^{\nu_j-\nu_k}-1}{\al_{jk}^2-1}
  =\vp_{\bnu}+\sum_{r=1}^{\frac12(\nu_j-\nu_k)-1}z_j^{-2r} z_k^{2r}\vp_{\bnu}\,,
\end{equation}
where the last sum only appears if $\nu_j-\nu_k>2$. In this case the multiindices $\bnu'$ of the
monomials in the summation symbol in Eq.~\eqref{firstterm} are of the form
\[
\bnu'=(\nu_1,\dots,\nu_j-2r,\dots,\nu_k+2r,\dots,\nu_N)\,,\qquad
r=1,\dots,\frac12(\nu_j-\nu_k)-1\,,
\]
and hence $\bnu'-\bnu\in(2\ZZ)^N$. Moreover, we have $0<\max\{\nu_j-2r,\nu_k+2r\}<\nu_j$ for all
$r=1,\dots,\frac12(\nu_j-\nu_k)-1$, so that $\bnu'\prec\bnu$. Thus, the first sum
in~\eqref{Jkphinu} contributes to $\la_{\bnu,k}$ the quantity
\begin{equation}
  2a\card\{j<k:\nu_j>\nu_k\}=2a\big(\ell(\nu_k)-1\big)\,.
  \label{contrib1}
\end{equation}
It may be likewise verified that the multiindices $\bnu'$ corresponding to the monomials arising
from the second sum in Eq.~\eqref{Jkphinu} either coincide with $\bnu$ or satisfy
$\bnu'\prec\bnu$, and that this sum yields the following contribution to $\la_{\bnu,k}$:
\begin{equation}
  2a\card\{j>k:\nu_j=\nu_k\}=2a\big(\ell(\nu_k)+\#(\nu_k)-k-1\big)\,.
  \label{contrib2}
\end{equation}
Consider, finally, the $j$-th term of the last sum in Eq.~\eqref{Jkphinu}. This term is equal to
$1$ when $\nu_j=\nu_k=0$, while for $\nu_j+\nu_k\ge2$ we have
\[
\frac{\be_{jk}^{2-\nu_j-\nu_k}-1}{\be^2_{jk}-1}=-\be_{jk}^{2-\nu_j-\nu_k}\,
\frac{\be_{jk}^{\nu_j+\nu_k-2}-1}{\be^2_{jk}-1}=-\sum_{r=1}^{\frac12(\nu_j+\nu_k)-1}\be_{jk}^{-2r}\,.
\]
Hence when $\nu_j+\nu_k\ge2$ the multiindices corresponding to the basis functions arising from
the last sum in Eq.~\eqref{Jkphinu} are of the form
\[
\bnu'=(\nu_1,\dots,\nu_j-2r,\dots,\nu_k-2r,\dots,\nu_N)\,,\qquad
r=1,\dots,\frac12(\nu_j+\nu_k)-1\,,
\]
and thus $\bnu'-\bnu\in(2\ZZ)^N$. Furthermore, since
\[
-\nu_k+2\le \nu_j'\le\nu_j-2\,,\qquad -\nu_j+2\le \nu_k'\le\nu_k-2\,,
\]
we have $\max\{|\nu_j'|,|\nu_k'|\}<\max\{\nu_j,\nu_k\}$, so that again $\bnu'\prec\bnu$. The
contribution of the last sum in Eq.~\eqref{Jkphinu} to $\la_{\bnu,k}$ is therefore equal to
\begin{equation}
  2a\big(\#(0)-1\big)\de_{\nu_k,0}\,.
  \label{contrib3}
\end{equation}
Adding Eqs.~\eqref{contrib1}--\eqref{contrib3} to the first two terms in the RHS of
Eq.~\eqref{Jkphinu}, and taking into account that $l(0)+\#(0)=N+1$, we easily obtain
Eq.~\eqref{la} for $\la_{\bnu,k}$.\bigskip

Equation \eqref{Jifnstruct} does not hold in general if $\bnu$ does not belong to
$\big[\ZZ^N\big]$, so that Eq.~\eqref{la} does not yield the spectrum of the Dunkl operators
$J_k$. On the other hand, for the purposes of computing the spectrum of $H'$ we shall only need
the following weaker result: if $\bnu\in\ZZ^N$ is a multiindex all of whose components have the
same parity, then
\begin{equation}
  \label{Jifnarb}
  J_k\vp_{\bnu} =\sum_{\substack{\bnu'\in\ZZ^N\\ \bnu'-\bnu\in(2\ZZ)^N,\,[\bnu']\preceq[\bnu]}}
  \ga_{\bnu,k}^{\bnu'}\,\vp_{\bnu'}
\end{equation}
for some complex constants $\ga_{\bnu,k}^{\bnu'}$. Indeed, if $\bnu$ is as above, there is an
element $W$ belonging to the Weyl group of $BC_N$ type $\fW^{(\mathrm B)}$ such that
$\vp_{\bnu}=W\vp_{[\bnu]}$. Proceeding as in Ref.~\cite{FGGRZ03}, it is easy to show that
\[ [J_k,W]=\sum_{j=1}^{2^NN!}c_{jk}W_j\,,\qquad c_{jk}\in\RR\,,
\]
where $\fW^{(\mathrm B)}\equiv \{W_j:j=1,\dots,2^NN!\}$. We thus have
\[
J_k\vp_{\bnu}=W\big(J_k\vp_{[\bnu]}\big)+\sum_{j=1}^{2^NN!}c_{jk}W_j\vp_{[\bnu]}\,,
\]
and Eq.~\eqref{Jifnarb} follows immediately from~\eqref{Jifnstruct} and the fact that the partial
ordering $\prec$ and the parity of the components are invariant under the action of $\fW^{(\mathrm
  B)}$.\bigskip

From the previous results it is relatively straightforward to compute the spectrum of $H'$. More
precisely, we shall next show that the action of $H'$ on each Schauder subbasis
$\big\{\vp_\bn^{(\de)}\big\}_{\bn\in\ZZ^N}$, ordered as explained above, is upper triangular:
\begin{equation}\label{Hpvp}
  H'\vp_\bn^{(\de)}=E_\bn^{(\de)}\vp_\bn^{(\de)}+
  \sum_{\bnu'\prec\bnu}c^{(\de)}_{\bn'\bn}\vp_{\bn'}^{(\de)}\,,\qquad \nu_k\equiv 2n_k+\de\,,\;
  \nu'_k\equiv 2n'_k+\de\,,
\end{equation}
where $c^{(\de)}_{\bn'\bn}\in\CC$ and
\begin{equation}\label{Ende}
  E_\bn^{(\de)}=\sum_k\big([\bnu]_k+2a(N-k)\big)^2\,.
\end{equation}
Indeed, suppose first that the multiindex $\bnu$ in Eq.~\eqref{Hpvp} is nonnegative and
nonincreasing. Applying $J_k$ to both sides of Eq.~\eqref{Jifnstruct} we obtain
\[
J_k^2\vp_{\bnu}=\la^2_{\bnu,k}\vp_{\bnu}+\sum_{\substack{\bnu'-\bnu\in(2\ZZ)^N\\
    \bnu'\!\prec\,\bnu}}
\la_{\bnu,k}\,c_{\bnu,k}^{\bnu'}\,\vp_{\bnu'}+\sum_{\substack{\bnu'-\bnu\in(2\ZZ)^N\\
    \bnu'\!\prec\,\bnu}} c_{\bnu,k}^{\bnu'}\,J_k\vp_{\bnu'}\,.
\]
By Eq.~eq\ref{Jifnarb}, the last sum is a linear combination of basis functions
$\vp_{\bnu''}$ with $\bnu''\prec\bnu$ and $\bnu''-\bnu\in(2\ZZ)^N$. Therefore we can write
\[
J_k^2\vp_{\bnu}=\la^2_{\bnu,k}\vp_{\bnu}+\sum_{\substack{\bnu'-\bnu\in(2\ZZ)^N\\
    \bnu'\!\prec\,\bnu}} b_{\bnu,k}^{\bnu'}\,\vp_{\bnu'}\,,
\]
with $b_{\bnu,k}^{\bnu'}\in\CC$. Summing over $k$ and using the identity~\eqref{HpJs} we obtain
\begin{equation}
  \label{Hpphinu}
  H'\vp_\bnu=\Big(\sum_k\la_{\bnu,k}^2\Big)\vp_{\bnu}
  +\sum_{\substack{\bnu'-\bnu\in(2\ZZ)^N\\ \bnu'\!\prec\,\bnu}}
  \Big(\sum_kb_{\bnu,k}^{\bnu'}\Big)\vp_{\bnu'}\,.
\end{equation}
Suppose, next, that $\bnu\notin[\ZZ]^N$, and let $W\in\fW^{(\mathrm B)}$ be such that
$\vp_\bnu=W\vp_{[\bnu]}$. Since $H'$ is obtained from its $BC_N$ counterpart in Ref.~\cite{EFGR05}
by setting $b=b'=0$, and the latter operator commutes with all the elements of $\fW^{(\mathrm
  B)}$, it follows that $[H',W]=0$. By Eq.~\eqref{Hpphinu} applied to $\vp_{[\bnu]}$ we have
\[
H'\vp_\bnu=W\cdot H'\vp_{[\bnu]}=\Big(\sum_k\la_{[\bnu],k}^2\Big)\vp_{\bnu}
+\sum_{\substack{\bnu'-[\bnu]\in(2\ZZ)^N\\ \bnu'\!\prec\,[\bnu]}}
\Big(\sum_kb_{[\bnu],k}^{\bnu'}\Big)W\vp_{\bnu'}\,,
\]
which establishes~\eqref{Hpvp} with
\begin{equation}\label{Enla}
  E_\bn^{(\de)}=\sum_k\la_{[\bnu],k}^2\,.
\end{equation}
All that remains to be proven is Eq.~\eqref{Ende} for the eigenvalue $E_\bn^{(\de)}$. To this end,
let $\bp=[\bnu]$ and suppose that $p_{k-1}>p_k=\cdots=p_{k+r}>p_{k+r+1}\geq 0$, so that
$\ell(p_{k+j})=k$ and $\#(p_{k+j})=r+1$ for $j=0,\ldots,r$. Since
\[
\la_{\bp,k+j}=-p_{k+j}+2a(k+r-j-N)= -p_{k+r-j}+2a\big(k+r-j-N),\quad j=0,\ldots,r\,,
\]
we have
\begin{equation}
  \label{partialsum}
  \sum_{j=k}^{k+r}\la_{\bp,j}^2=\sum_{j=k}^{k+r}\big(p_j+2a(N-j)\big)^2\,.
\end{equation}
If, on the other hand, $p_{k-1}>p_k=\cdots=p_N=0$, the analog of Eq.~\eqref{partialsum} follows
directly from~\eqref{la}. This completes the proof of Eq.~\eqref{Ende}.

\subsection{Triangularization of $H$ and $\Hsc$}\label{sec.triang}

Using the results of the previous subsection, it is a straightforward matter to triangularize $H$
and $\Hsc$. Indeed, by the results in Subsection~\ref{subsec.ext}, this problem is equivalent to
the triangularization of the extensions $\tH$ and $\tHsc$ acting on their respective Hilbert
spaces $\fH\equiv\La\big(L^2(C)\otimes\Si\big)$ and $\fHsc\equiv\Lasc\big(L^2(C)\big)$

Let us start with the operator $\tH$. By Eqs.~\eqref{LaL2Si} and~\eqref{V0V1}, its Hilbert space
can be decomposed as the direct sum
\begin{equation}\label{fH}
  \fH=\bigoplus_{\substack{\ep=\pm\\ \de=0,1}}\La^\ep\big(\fH^{(\de)}\otimes\Si\big)\,.
\end{equation}
Let $f(\bx)$ be in the domain of $H'$, and let $\ket s\in\Si$ denote an arbitrary spin state.
Since $\tH$ coincides with $H'\otimes\id$ on $\fH$, and the latter operator commutes with $\La$
(indeed, it commutes with all the elements of $\fW^{(\mathrm B)}$, and hence of $\fW$), we have
\begin{equation}\label{tHHp}
  \tH\big[\La^\ep\big(f(\bx)\ket s\big)\big]=\La^\ep\big[\big(H'f(\bx)\big)\ket s\big]\,.
\end{equation}
As we saw in the previous subsection, $H'$ preserves the subspaces $\fH^{(\de)}$, which by the
latter equation implies that each of the four subspaces $\La^\ep\big(\fH^{(\de)}\otimes\Si\big)$
is invariant under $\tH$. We shall next verify that $\tH$ acts triangularly on a (non-orthogonal)
basis of $\La^\ep\big(\fH^{(\de)}\otimes\Si\big)$ of the form
\begin{equation}
  \label{psis}
  \psi_{\bn,\bs}^{\de,\ep}(\bx)=\La^\ep\big(\vp_\bn^{(\de)}(\bx)\ket\bs\big),
\end{equation}
ordered in such a way that $\psi_{\bn,\bs}^{\de,\ep}$ precedes $\psi_{\bn',\bs'}^{\de,\ep}$
whenever $\bnu\prec\bnu'$ (with $\bnu$ defined in~\eqref{bnu}, and similarly $\bnu'$). The spin
functions~\eqref{psis} are obviously a complete set (since the functions~\eqref{Dbasis} are a
basis of $L^2(C)$), but their linear independence is not assured unless we impose suitable
restrictions on the quantum numbers $\bn$ and $\bs$. More precisely,
the states~\eqref{psis} are a (non-orthogonal) basis of the Hilbert space
$\La^\ep\big(\fH^{(\de)}\otimes\Si\big)$ provided that the quantum numbers $\bn\in\ZZ^N$ and $\bs$
satisfy the following conditions:
\begin{subequations}\label{conds}
  \begin{align}
    &\text{\hphantom{ii}i)}\enspace n_1\geq\cdots\geq n_N\ge0\label{cond1}\\[2mm]
    &\text{\hphantom{i}ii)}\enspace s_i>s_j \text{ whenever } n_i=n_j \text{ and } i<j
    \label{cond2}\\[2mm]
    &\text{iii)\enspace If } \de=n_i=0 \text{ then } s_i\geq 0 \text{ for } \ep=1, \text{ while
    } s_i>0 \text{ for } \ep=-1.\label{cond3}
  \end{align}
\end{subequations}
Indeed, since
\[
\La^\ep(K_iS_i)=\ep\La^\ep\,,\qquad\La^\ep(K_{ij}S_{ij})=-\La^\ep\,,
\]
acting with suitable operators $K_iS_i$ and $K_{ij}S_{ij}$ on a spin function
$\vp_\bn^{(\de)}(\bx)\ket\bs$ with arbitrary $\bn\in\ZZ^N$ and $\bs$ one can easily show that the
corresponding state $\psi^{\de,\ep}_{\bn,\bs}$ is either zero or proportional to a
state~\eqref{psis} satisfying the above conditions. (Note, in this respect, that a
state~\eqref{psis} with $\de=n_i=s_i=0$ is symmetric under $(x_i,s_i)\to(-x_i,-s_i)$, and must
therefore vanish identically if $\ep=-1$.) This shows that the states~\eqref{psis} with
$\bn\in\ZZ^N$ and $\bs$ satisfying the above conditions are complete. Their linear independence is
easily checked.
\begin{remark}
  Conditions i)--ii) above are identical to the corresponding ones for the spin Calogero model of
  $D_N$ type studied in Ref.~\cite{BFG09}. As to the third one, the key difference is that in the
  present case the action of a coordinate sign reversing operator $K_i$ on a state
  $\vp_\bn^{(\de)}(\bx)\ket\bs$ no longer produces a state with the same quantum number $\bn$ (up
  to a constant factor) unless $\de=n_i=0$.
\end{remark}

\begin{remark}
  Since the functions $\vp_{\bn}^{(0)}$ with $\bn\in\ZZ^N$ form a basis of $L^2(C^{(B)})\subset
  L^2(C)$ (cf.~Ref.~\cite{EFGR05}), it follows that each subspace
  $\La^\ep\big(\fH^{(0)}\otimes\Si\big)$ properly contains the Hilbert space
  $\La^\ep\big(L^2(C^{(\mathrm B)})\otimes\Si\big)$ of the Sutherland spin model of $BC_N$
  type~\eqref{HSB} with chirality~$\ep$. Note, however, that the other sector
  $\La^+\big(\fH^{(1)}\otimes\Si\big)\oplus \La^-\big(\fH^{(1)}\otimes\Si\big)$ of $\fH$ has no
  counterpart in the $BC_N$ model. Thus, in contrast with the rational case~\cite{BFG09}, the
  Hilbert space of the $D_N$ model is larger than the direct sum of the Hilbert spaces of its
  $BC_N$ counterparts with both chiralities.
\end{remark}

Let us now examine the action of the operator $\tH$ on the basis of
$\La^\ep\big(\fH^{(\de)}\otimes\Si\big)$ given by Eqs.~\eqref{psis}-\eqref{conds}. It is easy to
show that
\begin{equation}
  \label{tHpsi}
  \tH\psi_{\bn,\bs}^{\de,\ep}=E_{\bn,\bs}^{\de,\ep}\psi_{\bn,\bs}^{\de,\ep}+
  \sum_{\substack{\bn',\bs'\\\bnu'\prec\bnu}}c^{\de\ep}_{\bn'\bs',\bn\bs}\,\psi_{\bn',\bs'}^{\de,\ep}\,,
\end{equation}
where $c^{\de\ep}_{\bn'\bs',\bn\bs}\in\CC$ and
\begin{equation}\label{Ensde}
  E_{\bn,\bs}^{\de,\ep}=\sum_k\big(2n_k+\de+2a(N-k)\big)^2\,.
\end{equation}
Indeed, from Eqs.~\eqref{Hpvp}-\eqref{Ende} and the identity~\eqref{tHHp} one immediately obtains
\begin{equation}
  \label{tHpsiprev}
  \tH\psi_{\bn,\bs}^{\de,\ep}=E_{\bn,\bs}^{\de,\ep}\psi_{\bn,\bs}^{\de,\ep}+
  \sum_{\bnu'\prec\bnu}c^{(\de)}_{\bn',\bn}\,\psi_{\bn',\bs}^{\de,\ep}\,.
\end{equation}
Although the quantum numbers $(\bn',\bs)$ appearing in the RHS of this equation do not necessarily
satisfy conditions~\eqref{conds}, there is an element $W\in\fW^{(\mathrm B)}$ such that
$(W\bn',W\bs)\equiv(\bn'',\bs'')$ do satisfy these conditions. Since the corresponding state
$\psi_{\bn'',\bs''}^{\de,\ep}$ differs from $\psi_{\bn',\bs}^{\de,\ep}$ by at most an overall
sign, and $[\bnu'']=[\bnu']\prec[\bnu]$ implies that $\bnu''\prec\bnu$, it is clear that we can
rewrite~\eqref{tHpsiprev} in the form~\eqref{tHpsi}.\bigskip

From Eq.~\eqref{tHpsi} it follows that the operator $\tH$ acts triangularly on the
(non-orthogonal) basis of $\La^\ep\big(\fH^{(\de)}\otimes\Si\big)$ in
Eqs.~\eqref{psis}-\eqref{conds}, ordered in such a way that $\psi_{\bn,\bs}^{\de,\ep}$ precedes
$\psi_{\bn',\bs'}^{\de,\ep}$ whenever $\bnu\prec\bnu'$. Moreover, the eigenvalues of the
restriction of $\tH$ to $\La^\ep\big(\fH^{(\de)}\otimes\nobreak\Si\big)$ are given by
Eq.~\eqref{Ensde}, with $\bn\in\ZZ^N$ and $\bs$ satisfying conditions~\eqref{conds}.
\begin{remark}\label{rem.degf}
  Since the numerical value of the eigenvalue~\eqref{Ensde} does not depend on $\bs$ or $\ep$, for
  any multiindex $\bn\in\big[\ZZ^N\big]$ the corresponding eigenvalue $E^{\de,\ep}_{\bn,\bs}$ has
  an associated degeneracy
  \begin{equation}
    \label{dnde}
    d_\bn^\de=d_{\bn}^{\de,+}+d_{\bn}^{\de,-}\,,
  \end{equation}
  where $d_{\bn}^{\de,\ep}$ is the number of basic spin states $\ket\bs$ satisfying
  conditions~\eqref{conds} for given $\ep$ and $\de$. These spin degeneracy factors will be
  computed below, when we discuss the partition function of this model.
\end{remark}

The spectrum of the scalar Hamiltonian $\tHsc$ can be computed in a similar way by exploiting the
fact that it coincides with $H'$ in its Hilbert space $\fHsc$, which by Eqs.~\eqref{LaL2}
and~\eqref{V0V1} is given by
\begin{equation}
  \label{Hilbertsc}
  \fHsc=\bigoplus_{\substack{\ep=\pm\\ \de=0,1}}\Lasc^\ep\big(\fH^{(\de)}\big)\,.
\end{equation}
Due to the identity
\[
\tHsc\big(\Lasc^\ep f(\bx)\big)=\Lasc^\ep\big(H'f(\bx)\big)\,,
\]
it is immediate to show that each of the four subspaces $\Lasc^\ep\big(\fH^{(\de)}\big)$ is
invariant under $\tHsc$. Just as in Section~\ref{sec.triang}), it can be verified that the
functions
\begin{equation}
  \label{psissc}
  \psi_{\bn}^{\de,\ep}(\bx)=\Lasc^\ep\big(\vp_\bn^{(\de)}(\bx)\big),
\end{equation}
where $\bn\in\ZZ^N$ and
\begin{equation}
  \label{nssc}
  n_1\ge\cdots\ge n_N\ge\frac12\,(1-\ep)(1-\de)\,,
\end{equation}
are a Schauder basis of $\Lasc^\ep\big(\fH^{(\de)}\big)$. (The last inequality is due to the fact
that if $\de=n_N=0$ the function $\psi_{\bn}^{0,-}$ is symmetric under $x_N\to-x_N$, and therefore
vanishes identically). Proceeding as above, it is straightforward to show that if we order the
basis~\eqref{psissc}-\eqref{nssc} so that $\psi_{\bn}^{\de,\ep}$ precedes $\psi_{\bn'}^{\de,\ep}$
whenever $\bnu\prec\bnu'$, the operator $\tHsc$ acts triangularly on it, with eigenvalues
$E_{\bn}^{\de,\ep}$ given by the RHS of Eq.~\eqref{Ensde}. Of course, due to the absence of
internal degrees of freedom, in this case the degeneracy factors $d_\bn^{\de,\ep}$ are equal to
one for all quantum numbers $\bn$, $\ep=\pm1$, and $\de=0,1$.

\begin{remark}\label{rem.ortho}
  It is well-known~\cite{BF97} that the eigenfunctions of the scalar $BC_N$ Sutherland model are
  of the form
  \begin{equation}
    \label{rhoB}
    \rho(\bx)\prod_i|\sin x_i|^b|\cos x_i|^{b'}\cdot P_{k}(\by)\,,
  \end{equation}
  where $P_{k}(\by)$ is a symmetric polynomial in the variables $y_i=\sin^2 x_i$ ($i=1,\dots,N$).
  The polynomials $P_{k}$, which can be regarded as multivariate generalizations of the classical
  Jacobi polynomials, are orthogonal in the hypercube $[0,1]^N$ with respect to the weight
  function
  \begin{equation}
    \label{wB}
    w^{(\mathrm B)}(\by)=\prod_{i<j}|y_i-y_j|^{2a}\cdot\prod_iy_i^{b-\frac12}(1-y_i)^{b'-\frac12}
  \end{equation}
  (cf.~Eq.~(2.17) of Ref.~\cite{BF97}). In our case, from the identities
  \[
  \Lasc^+\,\e^{\iu\sum_k\nu_kx_k}=\Lasc\prod_k\cos(\nu_k x_k)\,,\qquad
  \Lasc^-\,\e^{\iu\sum_k\nu_kx_k}=\iu^N\Lasc\prod_k\sin(\nu_k x_k)\,,\qquad
  \]
  it is straightforward to show that the (orthonormalized) eigenfunctions of $\tHsc$ in each of
  the invariant subspaces $\Lasc^\ep\big(\fH^{(\de)}\big)$ are of the form
  \begin{equation}\label{Pndeep}
    \rho(\bx)\prod_i|\sin 2x_i|^{\frac{1-\ep}2}|\cos x_i|^{\de\ep}\cdot P_{k}^{\de,\ep}(\by)\,,
  \end{equation}
  where $P_{k}^{\de,\ep}(\by)$ is a polynomial in the variables $y_i=\sin^2 x_i$ symmetric under
  permutations. From the discussion in Section~\ref{subsec.ext} it follows that the restrictions
  of the functions~\eqref{Pndeep} to the open set $A$ are a complete set of eigenfunctions of the
  scalar Sutherland model of $D_N$ type $\Hsc$. They are also orthogonal in the latter set, on
  account of their symmetry under coordinate permutations and sign changes. This is easily seen to
  imply that the polynomials $P^{\de,\ep}_k$ are orthogonal in the hypercube $[0,1]^N$ with
  respect to the weight
  \begin{equation}\label{wdeep}
    w^{\de,\ep}(\by)=\prod_{i<j}|y_i-y_j|^{2a}\cdot\prod_iy_i^{-\frac\ep 2}(1-y_i)^{\ep(\de-\frac12)}\,.
  \end{equation}
  In view of Eqs.~\eqref{rhoB}-\eqref{wB} and~\eqref{Pndeep}-\eqref{wdeep}, it is clear that the
  three orthogonal polynomial families $\big\{P_k^{\de,\ep}:k\in\NN\big\}$ with
  $(\de,\ep)=(0,-1),(1,\pm1)$ are not limiting cases of the multivariate Jacobi polynomials studied
  by Baker and Forrester~\cite{BF97}. The analysis of the properties of these new orthogonal
  polynomials, and their relations with their $BC_N$ counterparts, could lead to interesting new
  developments in the field of multivariate orthogonal polynomials.
\end{remark}

\section{Partition function of the spin chain}\label{sec.pf}
The purpose of this section is to evaluate in closed form the partition function of the
Haldane--Shastry spin chain of $D_N$ type~\eqref{cH}. Our starting point is the freezing trick
relation~\eqref{frtr}, which can be equivalently written as
\begin{equation}
  \cE_j=\lim_{a\to\infty}\frac{E_{ij}-\Esc_i}{4a}\,.
  \label{frtrlim}
\end{equation}
This formula expresses each eigenvalue $\cE_j$ of the chain~\eqref{cH} in terms of a certain
eigenvalue $E_{ij}$ of the spin Sutherland model of $D_N$ type~\eqref{H} and a corresponding
eigenvalue $\Esc_i$ of the scalar model~\eqref{Hsc}. In practice, the fact that the eigenvalues
$E_{ij}$ and $\Esc_i$ are obviously not independent makes it impossible to use Eq.~\eqref{frtrlim}
to completely determine the spectrum of the chain~\eqref{cH} in terms of the spectra of the
Hamiltonians $H$ and $\Hsc$ computed in the previous section (cf.~Eq.~\eqref{Ensde}). The key idea
behind the freezing trick method introduced by Polychronakos~\cite{Po94} is to use
Eq.~\eqref{frtrlim}, or rather the equivalent relation~\eqref{frtr}, to directly compute the
chain's partition function. Indeed, the latter equation immediately yields the identity
\begin{equation}
  \label{ZZZ}
  \cZ(T)=\lim_{a\to\infty}\frac{Z(4aT)}{\Zsc(4aT)}\,,
\end{equation}
expressing the chain's partition function $\cZ$ in terms of the partition functions $Z$ and $\Zsc$
of the Hamiltonians $H$ and $\Hsc$.
\begin{remark}\label{rem.int}
  Equation~\eqref{frtrlim} can be used to obtain nontrivial qualitative information on the
  spectrum of the chain~\eqref{cH}. For instance, from the fact that the numerical values of the
  energies of both Hamiltonians $H$ and $\Hsc$ are given by the RHS of Eq.~\eqref{Ensde} and
  Eq.~\eqref{frtrlim} it easily follows that all the energies of the spin chain~\eqref{cH} are
  integers.
\end{remark}

In the rest of this section, we shall compute the large $a$ limits of $Z(4aT)$ and $\Zsc(4aT)$
using Eq.~\eqref{Ensde} for the spectrum of $H$ and $\Hsc$, thereby obtaining an exact expression
for $\cZ$ via Eq.~\eqref{ZZZ}. Before doing so, it is convenient to subtract from the spectra of
$H$ and $\Hsc$ the constant term
\[
E_0 = 4a^2\sum_k(N-k)^2 = \frac23\, a^2N(N-1)(2N-1)\,,
\]
which is of course irrelevant for the purposes of computing $\cZ$. The rationale behind this
normalization is the fact that, by Eq.~\eqref{Ensde}, the eigenvalues of $H$ and $\Hsc$ become
$O(a)$ for $a\to\infty$, so that the limits of $Z(4aT)$ and $\Zsc(4aT)$ exist separately.

Let us start with the partition function of Hamiltonian $H$ of the $D_N$-type spin Sutherland
model~\eqref{H}. With the normalization of the energies explained above, the spectrum of this
model satisfies
\begin{equation}
  \label{specHas}
  E_{\bn,\bs}^{\de,\ep} = 4a\sum_k(2n_k+\de)(N-k)+O(1)\,,
\end{equation}
and hence its partition function is given by
\begin{equation}
  \label{ZaT}
  \lim_{a\to\infty}Z(4aT)=\sum_{\substack{n_1\ge\cdots\ge n_N\ge0\\\ep=\pm,\ \de=0,1}}
  d_\bn^{\de,\ep}\,q^{\sum\limits_i(2n_i+\de)(N-i)}\,,\qquad q\equiv\e^{-1/(k_{\mathrm B}T)}\,.
\end{equation}
As mentioned in Remark~\ref{rem.degf}, the degeneracy factor $d_\bn^{\de,\ep}$ is equal to the
number of spin states $\ket{\bs}$ satisfying conditions~\eqref{conds} for given $\ep=\pm1$ and
$\de=0,1$. Writing the quantum number $\bn$ in the form
\begin{equation}
  \bn=\big(\overbrace{\vphantom{1}p_1,\dots,p_1}^{k_1},\dots,
  \overbrace{\vphantom{1}p_r,\dots,p_r}^{k_r}\big),\qquad p_1>\cdots>p_r\geq0,
  \label{neven}
\end{equation}
and using conditions~\eqref{cond2} and~\eqref{cond3} we have
\begin{equation}
  \label{dnepsde}
  d_\bn^{\de,\ep}=
  \begin{cases}\displaystyle
    \binom{m_\ep}{k_r}\prod\limits_{i=1}^{r-1}\binom{m}{k_i}\,,\quad& \de=p_r=0\,;\\[5mm]
    \displaystyle \hfill\prod\limits_{i=1}^r\binom{m}{k_i}\,,\hfill& \text{otherwise,}
  \end{cases}
\end{equation}
with
\begin{equation}
  \label{mep}
  m_\ep=\frac12\,\big(m+\ep\pi(m)\big)\,,\qquad \pi(m)\equiv m \pmod 2\,.
\end{equation}
Let us now define
\[
Z^{(\de)}(T)\equiv\sum_{\substack{n_1\ge\cdots\ge n_N\ge0\\\ep=\pm}}
d_\bn^{\de,\ep}\,q^{\sum\limits_i(2n_i+\de)(N-i)}\,,
\]
so that
\begin{equation}
  \label{Z0Z1}
  \lim_{a\to\infty}Z(4aT)=Z^{(0)}(T)+Z^{(1)}(T)\,.
\end{equation}
The function $Z^{(0)}(T)$ can be easily expressed in terms of the partition functions
$Z_\pm^{(\mathrm B)}$ of two spin Sutherland models of $BC_N$ type~\eqref{HSB} with opposite
chiralities $\ep=\pm$. Indeed, when $\de=0$ Eq.~\eqref{dnepsde} coincides with Eq.~(51) in
Ref.~\cite{EFGR05} for the degeneracy factor of the $BC_N$-type spin Sutherland model with
chirality $\ep$. Likewise, Eq.~\eqref{Ensde} with $\de=0$ is obtained from the analogous formula
for the energies of the $BC_N$ Hamiltonian~\eqref{HSB} in Eq.~(24) of the latter reference by
setting $\be=\be'=0$, and the same is true for Eq.~\eqref{specHas}. We thus have
\begin{equation}
  \label{Z0ZB}
  Z^{(0)}(T)=
  \lim_{a\to\infty}\Big(Z_+^{(\mathrm B)}(4aT)+Z_-^{(\mathrm B)}(4aT)\Big)\Big|_{\be=\be'=0}\,.
\end{equation}
Using Eq.~(53) from Ref.~\cite{EFGR05} we obtain the explicit expression
\begin{multline}
  \label{Z0}
  Z^{(0)}(T)=\sum_{(k_1,\ldots,k_r)\in\cP_N}\bigg\{
  \bigg[\binom{m_+}{k_r}+\binom{m_-}{k_r}\\
  +2\binom{m}{k_r} \,\frac{q^{K_r}}{1-q^{K_r}}\bigg]
  \prod_{i=1}^{r-1}\bigg[\binom{m}{k_i}\,\frac{q^{K_i}}{1-q^{K_i}}\bigg]\bigg\}\,,
\end{multline}
where $\cP_N$ is the set of partitions of the integer $N$ (taking order into account), and
\begin{equation}
  \label{Ki}
  K_i=\bar k_i\big(2N-1-\bar k_i\big)\,,\qquad \bar k_i\equiv\sum_{j=1}^ik_i\,.
\end{equation}
Note that, since $k_1+\cdots+k_r=N$, the integers $\bar k_i$ are in the range $1,\dots,N$.

On the other hand, from Eq.~\eqref{dnepsde} it easily follows that
\[
Z^{(1)}(T)= 2q^{\frac12N(N-1)}\kern-1em\sum_{n_1\ge\cdots\ge n_N\ge0}\prod_{i=1}^r\binom{m}{k_i}
\cdot q^{\sum\limits_j2n_j(N-j)}.
\]
Proceeding as in Ref.~\cite{EFGR05} we easily obtain
\begin{equation}
  \label{Z1}
  Z^{(1)}(T)=2q^{\frac12N(N-1)}\sum_{(k_1,\ldots,k_r)\in\cP_N}(1-q^{K_r})^{-1}
  \prod_{i=1}^r\binom m{k_i}\cdot
  \prod_{i=1}^{r-1}\frac{q^{K_i}}{1-q^{K_i}}\,.
\end{equation}

The partition function $Z_{\mathrm{sc}}(4aT)$ of the scalar Hamiltonian~\eqref{Hsc} is also easily
evaluated in the limit $a\to\infty$, since in this limit its spectrum (with the normalization
discussed above) is still given by the RHS\ of Eq.~\eqref{specHas}. Using Eq.~\eqref{nssc}, and
taking into account that in this case $d_\bn^{\de,\ep}=1$, we have
\begin{align*}
  \lim_{a\to\infty} Z_{\mathrm{sc}}(4aT)&=2\sum_{n_1\ge\cdots\ge n_N\ge0}
  q^{\sum\limits_i(2n_i+1)(N-i)}
  +\sum_{n_1\ge\cdots\ge n_N\ge0}q^{\sum\limits_i 2n_i(N-i)}\\
  &\qquad{}+\sum_{n_1\ge\cdots\ge
    n_N>0}q^{\sum\limits_i2n_i(N-i)}\\
  &=\big(2q^{\frac12 N(N-1)}+1\big)\sum_{n_1\ge\cdots\ge n_N\ge0}q^{\sum\limits_i 2n_i(N-i)}
  +\sum_{n_1\ge\cdots\ge n_N\ge0}q^{\sum\limits_i2(n_i+1)(N-i)}\\
  &=\big(1+q^{\frac12 N(N-1)}\big)^2\kern-7pt\sum_{n_1\ge\cdots\ge n_N\ge0}q^{\sum\limits_i
    2n_i(N-i)}.
\end{align*}
The last sum is easily recognized as the $a\to\infty$ limit of the partition function
$Z_{\mathrm{sc}}^{(\mathrm B)}(4aT)$ of the scalar Sutherland Hamiltonian of $BC_N$ type
\[
H^{(\mathrm B)}_{\mathrm{sc}}\equiv H^{(\mathrm B)}\Big|_{S_{ij}\to1,S_i\to1}
\]
with $\be=\be'=0$. Using Eq.~(49) in Ref.~\cite{EFGR05} we thus obtain
\begin{multline}
  \label{Zsc}
  \lim_{a\to\infty}Z_{\mathrm{sc}}(4aT)
  =\big(1+q^{\frac12N(N-1)}\big)^2\,\lim_{a\to\infty}Z_{\mathrm{sc}}^{(\mathrm
    B)}(4aT)\big|_{\be=\be'=0}\\
  =\big(1+q^{\frac12N(N-1)}\big)^2\prod_i\big(1-q^{i(2N-1-i)}\big)^{-1}\,.
\end{multline}

The partition function of the Haldane--Shastry spin chain of $D_N$ type~\eqref{cH} is easily
computed by inserting Eqs.~\eqref{Z0Z1}, \eqref{Z0}, \eqref{Z1} and~\eqref{Zsc} into the freezing
trick identity~\eqref{ZZZ}. In order to simplify the resulting expression, we define $N-r$
integers $\bar k_1'<\cdots<\bar k_{N-r}'$ in the range $1,\dots,N-1$ by
\[
\big\{\bar k_1',\dots,\bar k_{N-r}'\big\}=\big\{1,\dots,N-1\big\}-\{\bar k_1,\dots,\bar
k_{r-1}\big\}\,,
\]
and set
\begin{equation}\label{Kpi}
  K'_i=\bar k_i'\,\big(2N-1-\bar k_i'\,\big)\,.
\end{equation}
Using this notation, the partition function of the chain~\eqref{cH} can be written as
\begin{multline}
  \label{cZ}
  \cZ(T)=\big(1+q^{\frac12 N(N-1)}\big)^{-2}\kern-12pt\sum_{(k_1,\ldots,k_r)\in\cP_N}
  \prod_{i=1}^{r-1}\binom{m}{k_i}\cdot q^{\sum\limits_{i=1}^{r-1}K_i}
  \bigg\{2\binom{m}{k_r}\big(q^{K_r}+q^{\frac12 N(N-1)}\big)\\
  +\left[\binom{m_+}{k_r}+\binom{m_-}{k_r}\right] \big(1-q^{K_r}\big)\bigg\}\,
  \prod_{i=1}^{N-r}\big(1-q^{K'_i}\big)\,.
\end{multline}
Taking into account that $\bar k_r=\sum\limits_{i=1}^rk_i=N$, so that $K_r=N(N-1)$ by
Eq.~\eqref{Ki}, we finally obtain the more compact expression

\begin{multline}
  \label{cZsimp}
  \cZ(T)=\big(1+q^{\frac12 N(N-1)}\big)^{-1}\kern-12pt\sum_{(k_1,\ldots,k_r)\in\cP_N}
  \prod_{i=1}^{r-1}\binom{m}{k_i}\cdot
  \bigg\{2\binom{m}{k_r}q^{\frac12 N(N-1)}\\
  +\left[\binom{m_+}{k_r}+\binom{m_-}{k_r}\right]\big(1-q^{\frac12 N(N-1)}\big)\bigg\}\,
  q^{\sum\limits_{i=1}^{r-1}K_i}\prod_{i=1}^{N-r}\big(1-q^{K'_i}\big)\,.
\end{multline}
\begin{remark}\label{remZ}
  {}From~\cite[Eq.~(53)]{EFGR05} and Eq.~\eqref{cZ} we easily obtain the identity
  \[
  \cZ(T)=\big(1+q^{\frac12N(N-1)}\big)^{-2}\Big[ \Big(\cZ_+^{(\mathrm B)}(T)+\cZ_-^{(\mathrm
    B)}(T)\Big)\Big|_{\be=\be'=0} + 2q^{\frac12N(N-1)}\cQ_N(T)\Big]\,,
  \]
  where
  \begin{equation}
    \label{Ql}
    \cQ_l(T)=\sum_{(k_1,\dots,k_r)\in\cP_l}
    \prod_{i=1}^r\binom m{k_i}\ms\cdot\ms q^{\sum\limits_{i=1}^{r-1}K_i}\prod_{i=1}^{l-r}(1-q^{K'_i})
  \end{equation}
  and the integers $K_i$, $K'_i$ are defined by Eqs.~\eqref{Ki} and \eqref{Kpi} for all $l$. Thus,
  unlike what happens in the rational case (cf.~\cite[Eq.~(46)]{BFG09}), it does not seem possible
  to express in a simple way the partition function $\cZ(T)$ exclusively in terms of its $BC_N$
  counterparts $\cZ^{(\mathrm B)}_{\pm}$. Note also that the function $\cQ_N(T)$ has the same
  structure as the partition function of the ordinary ($A_N$-type) Haldane--Shastry
  chain~\cite{FG05}, the only difference being the ``dispersion relation'' defining the quantities
  $K_i$ and $K'_i$ in terms of $\bar k_i$ and $\bar k_i'$.
\end{remark}

As mentioned in Remark~\ref{rem.int}, the eigenvalues of the spin chain~\eqref{cH} are integers,
and they are nonnegative on account of the nonnegative character of the operators $1+S_{ij}$ and
$1+\tS_{ij}$. Thus, the partition function~$\cZ(T)$ should be a polynomial in $q$, a fact which is
not apparent from Eq.~\eqref{cZsimp}. In order to ascertain this fact, consider first a partition
$(k_1,\dots,k_r)\in\cP_N$ with $k_r=1$. In this case the term in curly brackets in
Eq.~\eqref{cZsimp} reduces to $m\big(1+q^{N(N-1)/2}\big)$, and $\bar k_{r-1}=N-k_r=N-1$ implies
that
\[
\big\{\bar k_1',\dots,\bar k_{N-r}'\big\}=\big\{1,\dots,N-2\big\}-\big\{\bar k_1,\dots,\bar
k_{r-2}\big\}\,.
\]
Hence the contribution to~$\cZ(T)$ of the partitions with $k_r=1$ is  given by
$m\,q^{N(N-1)}\cQ_{N-1}(T)$. Consider next a partition $(k_1,\dots,k_r)\in\cP_N$ such that
$k_r\equiv l>1$. In this case $(k_1,\dots,k_{r-1})$ is a partition of $N-l$, and $\bar
k_{r-1}=N-l$ implies that
\begin{equation}\label{kpi}
  \bar k_{N-j-r+1}'=N-j\,,\qquad j=1,\dots,l-1\,,
\end{equation}
and hence
\[
K'_{N-j-r+1}=(N-j)(N+j-1)\,,\qquad j=1,\dots,l-1\,.
\]
Note, in particular, that $K'_{N-r}=N(N-1)$, so that
\[
\big(1+q^{\frac12 N(N-1)}\big)^{-1}\big(1-q^{K_{N-r}'}\big)=1-q^{\frac12 N(N-1)}\,.
\]
Taking into account that, by Eq.~\eqref{kpi},
\[
\big\{\bar k_1',\dots,\bar k_{N-l-r+1}'\big\}=\big\{1,\dots,N-l-1\big\}-\big\{\bar k_1,\dots,\bar
k_{r-2}\big\}\,,
\]
it is immediate to verify that the contribution to $\cZ(T)$ of the partitions with $k_r=l\ge2$ is
given by
\begin{multline*}
  \big(1-q^{\frac12 N(N-1)}\big)q^{(N-l)(N+l-1)}
  \prod_{i=1}^{l-2}\big(1-q^{(N-i-1)(N+i)}\big)\\
  \times\bigg\{2\binom{m}{l}q^{\frac12
    N(N-1)}+\left[\binom{m_+}{l}+\binom{m_-}{l}\right]\big(1-q^{\frac12
    N(N-1)}\big)\bigg\}\cQ_{N-l}(T)\,,
\end{multline*}
with $\cQ_0\equiv 1$. Thus the partition function~\eqref{cZsimp} can be expressed as\footnote{Note
  that the terms with $l>m$ in the previous equation vanish identically due to the binomial
  coefficients. This is in fact a consequence of conditions~\eqref{cond2} and~\eqref{cond3},
  cf.~Eq.~\eqref{dnepsde}.}
\begin{align}
  \label{cZpoly}
  \cZ(T)&=m\,q^{N(N-1)}\cQ_{N-1}(T)\notag\\
  &+\big(1-q^{\frac12 N(N-1)}\big)\sum_{l=2}^{\min(m,N)}q^{(N-l)(N+l-1)}
  \prod_{i=1}^{l-2}\big(1-q^{(N-i-1)(N+i)}\big)\notag\\
  &\times\bigg\{2\binom{m}{l}q^{\frac12
    N(N-1)}+\left[\binom{m_+}{l}+\binom{m_-}{l}\right]\big(1-q^{\frac12
    N(N-1)}\big)\bigg\}\cQ_{N-l}(T)\,,
\end{align}
where the RHS is clearly a polynomial in $q$ on account of Eq.~\eqref{Ql}. This remarkable formula
is one of the main results in the paper.
\begin{remark}\label{rem.motifs}
  {}From Eqs.~\eqref{cZsimp} or~\eqref{cZpoly} it is apparent that the partition function of the
  $D_N$ chain~\eqref{cH} has a much more complex structure than its $BC_N$ counterpart with
  $\be=\be'=0$, cf.~\cite[Eq.~(53)]{EFGR05}. In particular, while for the $BC_N$ chain one can
  find~\cite{BFGR10u} a description of the spectrum in terms of a suitable generalization of
  Haldane's \emph{motifs}~\cite{HHTBP92}, it is not clear how to implement such a description for
  the present chain. Note that, for HS chains of $A_{N}$ type, the existence of such a description
  is the key ingredient in the proof of the Gaussian character of their level density~\cite{EFG10}
  when the number of sites tends to infinity, which is of importance in the context of quantum
  chaos.
\end{remark}

\section{Concluding remarks}\label{sec.remarks}
As mentioned in the Introduction, reductions of the $BC_N$ Calogero and Sutherland models obtained
by setting suitable coupling constants to zero have been largely ignored in the extensive
literature devoted to these models. This is probably due to the fact that these reductions were
mostly regarded as trivial limits of the above models. In a previous paper~\cite{BFG09}, we showed
that this is not case by studying the $D_N$ reduction of the (spin) $BC_N$ Calogero model.
Moreover, the spin chain of Haldane--Shastry type associated with this reduction was also seen to
differ from its $BC_N$ counterpart even more markedly, essentially due to the nontrivial nature of
Polychronakos's ``freezing trick''. The aim of the present paper is to perform a comprehensive
study of the $D_N$ reduction of the $BC_N$ spin Sutherland model and its associated spin chain.

A significant part of the paper is devoted to the exact computation of the spectrum of the
dynamical spin model~\eqref{H} and its scalar version~\eqref{Hsc}. We have first provided a
rigorous proof of the equivalence of these models to their extended versions $\tH$ and $\tHsc$
defined on the Weyl-invariant configuration space $C$. The latter set, which turns out to be the
$N$-dimensional generalization~\eqref{Ceq} of a rhombic dodecahedron, is more complicated in
nature than its $BC_N$ counterpart (a hypercube). The motivation for constructing the extended
operators $\tH$ and $\tHsc$ is the fact that on their natural domains they essentially coincide
with the restriction of a simpler auxiliary operator $H'$, which can be expressed as a sum of
squares of a suitable set of Dunkl operators of $D_N$ type. In particular, from the spectrum of
$H'$ it is not difficult to deduce those of $\tH$ and $\tHsc$, and hence of $H$ and $\Hsc$. In
order to compute the spectrum of $H'$, we have next constructed an appropriate (non-orthogonal)
basis of the Hilbert space $L^2(C)$ where this operator acts. This is indeed the key difference
with the rational case, for which this step is trivial due to the fact that the (extended)
configuration spaces of both the $D_N$ and $BC_N$ models is $\RR^N$. Using a method similar to
that of Ref.~\cite{FGGRZ03}, we have shown that $H'$ acts triangularly on the above basis when
ordered appropriately. Finally, we have shown how to construct from the latter basis a
(non-orthogonal) basis of the Hilbert spaces of $\tH$ and $\tHsc$ where their action is also upper
triangular. In this way we have computed in closed form the spectra of the spin Sutherland model
of $D_N$ type~\eqref{H} and its scalar version~\eqref{Hsc}.

The second main result in the paper is the exact computation of the partition function of the
Haldane--Shastry spin chain of $D_N$ type~\eqref{cH} obtained from the spin dynamical
model~\eqref{H} by means of Polychronakos's freezing trick. The latter chain, as is apparent from
Eq.~\eqref{impurity}, cannot be obtained from its $BC_N$ counterpart by taking the limit
$(\be,\be')\to0$ due to the presence of an ``impurity'' term at both endpoints. Our starting point
is the fundamental relation~\eqref{ZZZ}, expressing the chain's partition function as the large
coupling constant limit of the quotient between the partition functions of the corresponding spin
dynamical model $H$ and its scalar version $\Hsc$. Using the above mentioned results for the
spectra of these models, we have been able to evaluate this limit, thereby obtaining
Eq.~\eqref{cZsimp} for the chain's partition function. In contrast with the rational case
(cf.~Remark~\ref{remZ}), the partition function is not expressed in a simple way in terms of its
$BC_N$ counterparts, since it also involves the partition function of the original (type $A$) HS
chain with a slightly different dispersion relation. We have further simplified Eq.~\eqref{cZsimp}
for the partition function, showing how to write it explicitly as a polynomial in
$q\equiv\e^{-1/(k_{\mathrm B}T)}$ (see Eq.~\eqref{cZpoly}), as should be the case for a finite
system. In fact, this simplified formula turns out to be quite efficient for the numerical
computation of the chain's spectrum, making it possible to perform a statistical analysis of the
spectrum when the number of particles becomes very large. It would be worthwhile to carry out such
a study, and compare its results with the corresponding ones for other spin chains of HS
type~\cite{BFGR08,BFGR08epl,BFGR09power,BB06,BB09,BFG09,FG05}.

The results of this paper suggest a number of further developments that we shall now discuss. In
the first place, we have shown that the $D_N$ reduction of the standard Sutherland model of $BC_N$
type gives rise to an interesting new solvable model that had been previously overlooked. In fact,
there are several additional reductions that could be considered, like e.g. those associated with
the $B_N$ and $C_N$ root systems, or even more general ones, like the $b=0$ reduction of the
Sutherland model~\eqref{HSB}. It could also be of interest to consider similar reductions of the
comparatively less studied hyperbolic Sutherland model of $BC_N$ type~\cite{FGGRZ03}. As we have
also shown in this work, these reductions can potentially yield unexpected results in other
fields, as for instance the remarkable tiling of $\RR^N$ with the $N$-dimensional generalization
of the rhombic dodecahedron uncovered in Section~\ref{sec.spectrum} (see Remark~\ref{rem.tiling}).

The work presented in this paper has direct implications in the field of multivariate orthogonal
polynomials. More precisely (see Remark~\ref{rem.ortho}), the eigenfunctions of the $D_N$
reduction of the scalar Sutherland model yield new families of multivariate orthogonal polynomials
that cannot be obtained as straightforward limits of the generalized Jacobi polynomials associated
with the $BC_N$ Sutherland model~\cite{BF97}. It is to be expected that the additional reductions
mentioned above could lead to similar new families of orthogonal polynomials.

An important aspect of spin chains of Haldane--Shastry type that has not been dealt with in this
paper is their integrability, which for the original HS chain of $A_{N}$ type was established by
constructing a transfer matrix satisfying the Yang--Baxter equation~\cite{BGHP93}. This matrix was
also used in the latter reference to derive the full Yangian symmetry algebra of this model, which
is ultimately responsible for the highly degenerate character of its spectrum. Moreover, the
representation theory of the Yangian is closely related to Haldane's elegant description of the
spectrum in terms of motifs~\cite{HHTBP92}. It is natural to inquire whether a similar
construction is possible for the $D_N$ chain of HS type studied in this paper. In fact, our
numerical calculations show that the spectrum of the $D_N$ chain is also highly degenerate, which
points to the existence of a large underlying symmetry algebra. The characterization of this
algebra, and its precise connection with the Yang--Baxter equation, is yet another open problem
motivated by the present work.

\section*{Acknowledgments}
This work was supported in part by the MICINN and the UCM--Banco Santander under grants
no.~FIS2008-00209 and~GR58/08-910556. The authors would also like to thank A.~Enciso for several
helpful discussions.


\end{document}